\documentclass[11pt,twocolumn]{aastex631}

\received{March 30, 2022}
\revised{April 11, 2022}
\accepted{April 12, 2022}
\submitjournal{ApJ}
\usepackage{ulem}

\newcommand{\pmra}{$\mu_\alpha \cos\delta$}
\newcommand{\pmdec}{$\mu_\delta$}

\shorttitle{Hierarchical Star Formation: Spatial and Kinematic Substructures}
\shortauthors{Pang et al.}

\defcitealias{pang2021a}{Paper~I}

\begin{document}

\title{3D Morphology of Open Clusters in the Solar Neighborhood with Gaia EDR\,3\,II: Hierarchical Star Formation Revealed by Spatial and Kinematic Substructures}

\author[0000-0003-3389-2263]{Xiaoying Pang}
    \affiliation{Department of Physics, Xi'an Jiaotong-Liverpool University, 111 Ren’ai Road, Dushu Lake Science and Education Innovation District, Suzhou 215123, Jiangsu Province, P. R. China}
    \email{Xiaoying.Pang@xjtlu.edu.cn}
    \affiliation{Shanghai Key Laboratory for Astrophysics, Shanghai Normal University, 
                100 Guilin Road, Shanghai 200234, P. R. China}

\author[0000-0003-4247-1401]{Shih-Yun Tang}
    \affiliation{Lowell Observatory, 1400 W. Mars Hill Road, Flagstaff, AZ 86001, USA}
    \affiliation{Department of Astronomy and Planetary Sciences, Northern Arizona University, Flagstaff, AZ 86011, USA}
    
\author{Yuqian Li}
    \affiliation{Department of Physics, Xi'an Jiaotong-Liverpool University, 111 Ren’ai Road, Dushu Lake Science and Education Innovation District, Suzhou 215123, Jiangsu Province, P. R. China}

\author[0000-0001-6980-2309]{Zeqiu Yu}
    \affiliation{Department of Physics, Xi'an Jiaotong-Liverpool University, 111 Ren’ai Road, Dushu Lake Science and Education Innovation District, Suzhou 215123, Jiangsu Province, P. R. China}
                
\author[0000-0001-8713-0366]{Long Wang}
    \affiliation{School of Physics and Astronomy, Sun Yat-sen University, Daxue Road, Zhuhai, 519082, China}

\author[0000-0001-6019-8467]{Jiayu Li}
    \affiliation{Department of Physics, Xi'an Jiaotong-Liverpool University, 111 Ren’ai Road, Dushu Lake Science and Education Innovation District, Suzhou 215123, Jiangsu Province, P. R. China}

\author[0000-0002-5740-5477]{Yezhang Li}
    \affiliation{Department of Physics, Xi'an Jiaotong-Liverpool University, 111 Ren’ai Road, Dushu Lake Science and Education Innovation District, Suzhou 215123, Jiangsu Province, P. R. China}

\author[0000-0002-7117-9533]{Yifan Wang}
    \affiliation{Department of Physics, Xi'an Jiaotong-Liverpool University, 111 Ren’ai Road, Dushu Lake Science and Education Innovation District, Suzhou 215123, Jiangsu Province, P. R. China}

\author[0000-0002-1243-8224]{Yanshu Wang}
    \affiliation{Department of Physics, Xi'an Jiaotong-Liverpool University, 111 Ren’ai Road, Dushu Lake Science and Education Innovation District, Suzhou 215123, Jiangsu Province, P. R. China}

\author[0000-0003-4976-6085]{Teng Zhang}
    \affiliation{Department of Physics, Xi'an Jiaotong-Liverpool University, 111 Ren’ai Road, Dushu Lake Science and Education Innovation District, Suzhou 215123, Jiangsu Province, P. R. China}

\author[0000-0003-3784-5245]{Mario Pasquato}
    \affiliation{Center for Astro, Particle and Planetary Physics (CAP$^3$), New York University Abu Dhabi}
    \affiliation{INFN- Sezione di Padova, Via Marzolo 8, I–35131 Padova, Italy}

\author[0000-0002-1805-0570]{M.B.N. Kouwenhoven}
    \affiliation{Department of Physics, Xi'an Jiaotong-Liverpool University, 111 Ren’ai Road, Dushu Lake Science and Education Innovation District, Suzhou 215123, Jiangsu Province, P. R. China}

\begin{abstract} 

We identify members of 65 open clusters in the solar neighborhood using the machine-learning algorithm \texttt{StarGO} based on Gaia EDR3 data. After adding members of twenty clusters from previous studies \citep{pang2021a,pang2021b,li2021} we obtain 85 clusters, and study their morphology and kinematics. We classify the substructures outside the tidal radius into four categories: filamentary (f1) and fractal (f2) for clusters $<100$\,Myr, and halo (h) and tidal-tail (t) for clusters $>100$\,Myr. The kinematical substructures of f1-type clusters are elongated; these resemble the disrupted cluster Group\,X. Kinematic tails are distinct in t-type clusters, especially Pleiades. We identify 29 hierarchical groups in four young regions (Alessi\,20, IC\,348, LP\,2373, LP\,2442); ten among these are new. The hierarchical groups form filament networks. Two regions (Alessi\,20, LP\,2373) exhibit global ``orthogonal'' expansion (stellar motion perpendicular to the filament), which  might cause complete dispersal. Infalling-like flows (stellar motion along the filament) are found in UBC\,31 and related hierarchical groups in the IC\,348 region. Stellar groups in the LP\,2442 region (LP\,2442 gp\,1--5) are spatially well-mixed but kinematically coherent. A merging process might be ongoing in the LP\,2442 subgroups. For younger systems ($\lesssim 30$\,Myr), the mean axis ratio, cluster mass and half-mass radius tend to increase with age values. These correlations between structural parameters may imply two dynamical processes occurring in the hierarchical formation scenario in young stellar groups: (1) filament dissolution and (2) sub-group mergers. 
\end{abstract}
\keywords{stars: evolution --- open clusters and associations: individual -- stars: kinematics and dynamics -- methods: statistical -- methods: numerical }


\section{Introduction}\label{sec:intro}

Stars are formed in molecular clouds where the environment is compacted with dense gas \citep{lada2003}. When the gravitational force inside molecular clouds overcomes the internal pressure after the cloud cools down, the gas starts to collapse and form stars. This contraction process can result in the formation of filaments which form a network feeding new materials into hubs at the intersections \citep{schneider2010,krause_physics_2020}. This mass transfer along the filaments is referred to as ``conveyor belt'' \citep{longmore_formation_2014,krumholz_star_2019}. Therefore, the interstellar medium inside the molecular cloud is highly hierarchically structured. Evidence from infrared and sub-millimeter observations shows that the pre- and protostellar cores' positions are strongly correlated with the hierarchy inside the clouds  \citep{elmegreen1996, testi_star_2000,gutermuth_spitzer_2009,konyves_census_2015-1,rathborne_cluster_2015,trevino-morales2019}. This observational evidence indicates that star formation is taking place across a continuous density distribution, from the densest hub regions to broader and lower-density filamentary networks \citep{trevino-morales2019}. The aforementioned observational results indicate a need for improvement of the ``monolithic cluster formation'' mechanism \citep{lada1984}. In the monolithic cluster formation scenario, star clusters were first born in the central densest region of molecular clouds. After a violent gas expulsion, stars in the dense clusters will expand and eventually disperse into the field, i.e., the ``infant mortality'' \citep{lada2003}.

To explain the star formation taking place in the hierarchical interstellar medium, \citet{kruijssen2012} developed the theoretical framework of ``hierarchical star formation''. In this paradigm, only star formation in the high-density region can produce gravitationally-bound star clusters. Stellar groups formed in low-density regions have filamentary substructures, and quickly become unbound after the residual gas is expelled. This model predicted that one-third of the field star population come from  bound clusters. Recent Gaia observations tend to support this scenario and suggest that only about 8--27\% of the stars in the solar neighborhood have originated from bound clusters \citep{anders_star_2021}.

Young stellar groups are the ideal targets to test the hierarchical formation theory. Their youth allows us to trace the morphological and kinematic features inherited from their parent molecular cloud \citep{vazquez-semadeni_hierarchical_2016,wright_kinematics_2018,ward_not_2020,krause_physics_2020}. Previous studies concluded that stars of OB associations are born in the low-density regions of molecular clouds, and that their filamentary substructures were formed in situ \citep{wright_cygnus_2016,wright_kinematics_2018,ward_not_2018,ward_not_2020}. The elongated filamentary substructures of stellar groups, the relic of filaments, are often associated with young ($\lesssim 100$\,Myr) open clusters \citep{jerabkova2019,cantat2019a,kounkel2019,beccari2020,pang2021b,kerr2021,wang2021} where significant kinematic substructures are also found \citep{wright_cygnus_2016,wright_kinematics_2018,pang2021b}. Young open clusters still have the morphological and kinematic substructures that support the nature of their unevolved dynamical status. Evidence of initial substructure in both morphology and kinematics is erased within 
a few global crossing times \citep{goodwin2004}. 

According to the ``conveyor belt'' mechanism in the framework of hierarchical formation, not only gas, but also stars are transferred into the hub regions. The in-falling flow brings smaller groups of stars along the filaments to the dense hub region \citep{vazquez-semadeni_hierarchical_2016,trevino-morales2019}. These in-falling subgroups will continue to merge when the dynamics of the inter-groups reach subvirial \citep{goodwin2004}. The morphology of the merged subgroups is then temporarily preserved. After sufficient relaxation in the same potential well of the hub region, sub-groups mix and lose their memory of the in-falling dynamics from the filament. However, if the dynamical state of the inter-groups transits from subvirial to supervirial during the in-falling process, sub-groups will just pass by each other. Numerical simulations have shown that star clusters form from the merging of sub-stellar groups at a dynamically subvirial stage \citep{mcmillan2007,allison2009,moeckel2009}. Moreover, multi-wavelength observations of clusters showing evidence of sub-group mergers \citep{kuhn_spatial_2015} support these simulation results. Nonetheless, more observations of young clusters ($<100$\,Myr) are required to fortify (strengthen) the hierarchical formation mechanism for star clusters. 

In our first paper of this series in studying the 3D morphology of open clusters in the Solar Neighborhood, \citet[][here after \citetalias{pang2021a}]{pang2021a}, we found three out of 13 target clusters already show signs of elongated filamentary substructures, i.e., young clusters NGC\,2232, NGC\,2547, and NGC\,2451B with an age of about 25--60 Myr. These filamentary substructure may be connected to a larger hierarchical network, e.g., NGC\,2232 is embedded in a string structure of a few hundred parsec long \citep{kounkel2019}. Our follow-up study \citep{pang2021b} investigated five hierarchical clustering groups in the Vela~OB2 region that showed a shell-like morphology related to kinematic substructures in the proper motion space. The significant expansion found in these hierarchical groups probably leads to a fate of dissolution. 
It is desirable to further search for more hierarchical groups and young clusters harboring spatial and kinematic substructures to understand the early formation process and evolution of star clusters. 
In this study, we investigate the morphology and kinematics of 85 open clusters (Table~\ref{tab:paramter_all}) located in the Solar Neighborhood primarily within 500\,pc based on Gaia Early Data Release 3 (EDR\,3) data \citep{gaia2021}. In our samples, 60 out of the 85 clusters are younger than 100\,Myr. We follow the method developed in \citetalias{pang2021a} to quantify stellar groups' spatial distribution and kinematic distribution in either three- or two-dimensional space.

This paper is organized as follows. In Section~\ref{sec:gaia_member}, we describe steps on the pre-processing of {\it Gaia} EDR\,3 data that serve as input for cluster member selection. We then present the algorithm, \texttt{StarGO}, used to determine cluster members and identify hierarchical clustering groups. The reliability of the cluster membership is verified in Section~\ref{sec:member_reliablity}. In Section~\ref{sec:dis_corr}, the distance correction for individual stars is carried out. 
The 3D morphologies and classification of the target open clusters are presented in Section~\ref{sec:3D_solar}. The analysis of their kinematic substructures is given in Section~\ref{sec:kin_feature}. The properties of the hierarchical groups is presented in Section~\ref{sec:hiearchy}. The morphology of the stars inside the tidal radius of each cluster is quantified in Section~\ref{sec:mor_tidal}. In Section~\ref{sec:3ddis}, we discuss the dependence of cluster morphology and its indication on star formation and evolution. Finally, we provide a summary in Section~\ref{sec:summary}.

\section{Cluster Member Identification and Verification}\label{sec:gaia_member}

To increase the young open cluster sample size in \citetalias{pang2021a}, we apply the search for hierarchical stellar grouping in fifty open cluster regions in the solar neighborhood (up to a distance of about 650\,pc). Unlike \citetalias{pang2021a}, where we only focus on studying the 3D morphology of the 13 clusters, and ignore stellar groups in the nearby sky region, we include all nearby closely-related structures of the target region in this study, e.g., hierarchical clustering groups or neighboring clusters.

\subsection{Gaia EDR\,3 Data Preprocessing}\label{sec:preprocess}

The data pre-processing procedures prior to the member selection (Section~\ref{sec:stargo}) are similar to our previous studies \citep{tang2019,pang2020,pang2021a,pang2021b}. In short, we query the Gaia EDR\,3 database for the regions around fifty target regions using a spherical spatial cut in the Cartesian coordinates. A target region can contain either just one open cluster or several clusters and/or stellar groups at the same time. The radius of the spherical cut is either 100\,pc or 150\,pc from the cluster center, depending on the size of the stellar groups present in the region. We then apply a proper motion cut for each cluster region based on a 2D proper motions (PMs) density map (e.g., Figure~1 in \citetalias{pang2021a}). The PM cuts are done circularly to include as many potential members in the target region as possible. The average spatial coordinates and proper motions of clusters used for the above selection are taken from \citet{liu2019} and \citet{Cantat-Gaudin2020}. The data quality cut for the Gaia data follows the instruction in the Appendix~C of \citet{lindegren2018} to select stars with parallaxes and photometric measurements less than 10 percent uncertainty.

\subsection{Membership Determination}\label{sec:stargo}

We use \texttt{StarGO}\footnote{\url{https://github.com/zyuan-astro/StarGO-OC}} \citep{yuan2018}, an unsupervised machine learning software based on the Self-Organizing-Map (SOM) algorithm, to select cluster members in data after the PM cut. \texttt{StarGO} has been successful in membership identification in both open clusters \citep{tang2019,pang2020,pang2021a,pang2021b} and stellar streams \citep{yuan2020a,yuan2020b}. Detailed descriptions about \texttt{StarGO} were given in the studies mentioned above; thus, here, we only give a brief summary for \texttt{StarGO}. The SOM algorithm starts with building a 2D neuron-map with each neuron initially assigned with five random weight vectors. The dimension of the random weight vectors matches the number of the input parameters, which are $X$, $Y$, $Z$, \pmra, and \pmdec{} for this study. We adopt a network of either 100$\times$100 or 150$\times$150 neurons depending on the number of input stars. During the learning process, target stars are fed to the 2D neuron-map one by one so that the weight vectors of each neuron can be updated to become closer to the input vectors of a given star. One iteration of the training cycle is finished after the neurons are trained by every star in the sky region. For the weight vectors in all neurons to converge, the training cycle is set to iterate 400 times. After the training process of the 2D neuron-map, each star will be assigned to a neuron where the five parameters of the star are closest to the neuron's weight vectors.

A cluster of stellar objects will appear as a grouping in the 5-D input parameter space. In terms of the trained 2D neuron-map, this grouping will show as a local minimum of the $u$ values\,---\,the difference of weight vectors between adjacent neurons. The final cluster members are stars associated with neurons below a certain $u$ value which is determined by a $\sim$5\% field star contamination rate \citep[e.g., Figure~1 panels (b) and (c) in ][]{pang2020}. The estimation of the field star contamination rate is computed with stars from the mock Gaia EDR\,3 catalog \citep{rybizki2020}. After applying the same selection criteria described in Section~\ref{sec:preprocess} to the mock stars, those remaining mock stars attached to the neurons same as the cluster members are considered as false positive, i.e., field star contaminants. Assuming the smooth Milky Way population has the same properties as the mock catalog, the contamination rate of each identified group can be estimated as the number of the field star contaminants over the selected cluster members \citep[more details in][Section~2.2]{pang2020}.

For regions showing significant substructures in the 2D neural network, a sign of hierarchical clustering, we apply a top-down subgroup selection method developed in \citet[][detailed in Section~2.2 and Figure~1 (c)]{pang2021b} to further disentangle them. In this top-down subgroup selection method, the top-level clustering is the 2D neural patch regions associated with $u$-values corresponding to 5\% contamination. To reveal the hierarchical substructures, we decrease the threshold value of $u$ until the contamination rate reaches 1--2\%. The remaining patches are considered as the core regions of each subgroup. We then associate neurons in the top-level structures to these cores by calculating the minimum difference in weight vectors between a given neuron to each core patch.

An additional member selection process is carried out in the color-magnitude diagram. Similar to Section~2.3 in \citet{pang2020}, we first fit the PARSEC isochrone \citep[version 1.2S,][]{bressan2012,chen2015} to all clusters and stellar groups by eye with reddening and metallicity values adopted from literature \citep{Villanova2009, zari2017,gaia&babusiaux2018, gaia2018VODC, Bossini2019, Zhang2019, carrera2019, Roeser2019, casali2020, pang2021a}. Stars below the best-fit isochrone or an isochrone with older age for young clusters ($<10$\,Myr) were then removed. The mass of each member star is then estimated from the best-fit isochrone using the k–D tree method \citep{mcmillan2007} to find the nearest point on the isochrone.

Finally, there are 65 clusters and stellar groups, with a total of 22,347 members, identified in the original fifty target regions. We provide basic information of these 65 clusters in Table~\ref{tab:paramter_all} obtained in this study and their complete member lists in Table~\ref{tab:memberlist}. Among the identified 65 clusters, 36 cross-match with clusters in \citet{liu2019} and 52 cross-match with clusters in \citet{Cantat-Gaudin2020}. On average, the matched stars account for 70--80\% of members in \citet{liu2019} and \citet{Cantat-Gaudin2020}. In general,
we double the number of member stars in these matched clusters compare to previous studies \citep{liu2019,Cantat-Gaudin2020}.
Among the 29 hierarchical groups identified in our samples, one-third (ten groups) are newly-discovered; these are located in regions of Alessi\,20, IC\,348, LP\,2373 and LP\,2442.

To further enlarge our sample size for a more robust statistical study on the 3D morphology and kinematic structures of nearby clusters and groups, we add 20 additional clusters with members from our previous studies. Twelve clusters are from \citetalias{pang2021a}: IC\,2391, IC\,2602, IC\,4665, NGC\,2422, NGC\,2516, NGC\,2547, NGC\,6633, NGC\,6774, NGC\,2451A, NGC\,2232, Blanco\,1, and Coma Berenices, seven stellar groups (clusters) from the Vela~OB2 region in \citet[][Huluwa 1--5, Collinder\,135, and UBC\,7]{pang2021b}, and the final one, the Pleiades open cluster, from \citet{li2021}. Although a list of members of NGC\,2415B was already provided in \citetalias{pang2021a}, we re-select its member in this study (increased by 10\%) in a larger spherical space to better probe its extended substructures. Parameters of the above-mentioned 20 extra clusters are obtained from previous studies and included in Table~\ref{tab:paramter_all}, except for the bound mass fraction ($f_m$) and the morphological type (flag) that are determined in this work. Therefore, we use 85 open clusters for analysis in this study. Although some of the target cluster members come from other studies, they were all obtained with the same procedure of member identification in the current study, and therefore form a homogeneous data set that is ideal for statistical analysis. 

\startlongtable
\begin{deluxetable*}{L RR R RRR RRRR RRR L}
	\tablecaption{Parameters of 85 open clusters.
	\label{tab:paramter_all}}
    \tabletypesize{\footnotesize}
	\tablehead{	    
	    \colhead{Cluster}   & 
	    \colhead{R.A.}      & \colhead{Decl.}   &
	    \colhead{Dist.}     & \colhead{RV}      & 
		\colhead{\pmra}     & \colhead{\pmdec}  & \colhead{Age} & 
		\colhead{$M_{cl}$}  & 
		\colhead{$r_{\rm h}$} & \colhead{$r_{\rm t}$} & 
		\colhead{N}         & \colhead{$f_{\rm m}$}   &
        \colhead{flag}
		\\
		\colhead{}          & 
		\multicolumn{2}{c}{(deg)}   & 
		\colhead{(pc)}      & \colhead{(km~s$^{-1}$)}  & 
		\multicolumn{2}{c}{(mas yr$^{-1}$)}     & \colhead{(Myr)}   & 
		\colhead{($M_\odot$)}   & 
		\multicolumn{2}{c}{(pc)}    & 
		\colhead{}          & \colhead{}    & \colhead{} 
		\\
	    \cline{2-3} \cline{6-7} \cline{10-11}
	    \colhead{(1)}   & 
	    \colhead{(2)}   & \colhead{(3)} & 
	    \colhead{(4)}   & \colhead{(5)} &
	    \colhead{(6)}   & \colhead{(7)} & \colhead{(8)} & 
	    \colhead{(9)}   & 
	    \colhead{(10)}  & \colhead{(11)} & 
	    \colhead{(12)}  & \colhead{(13)} & 
	    \colhead{(14)}
		}
	\startdata
	\rm Alessi\,3	&	109.139147	&	-46.074168	&	277.3	&	1.22	&	-9.804	&	11.945	&	631	&	124.6	&	6.8	&	7.0	&	165	&	0.52	&	\rm t	\\
\rm Alessi\,5	&	160.927572	&	-61.087405	&	397.7	&	11.13	&	-15.354	&	2.535	&	52	&	244.0	&	4.4	&	8.7	&	279	&	0.89	&		\\
\rm Alessi\,9	&	265.946512	&	-47.070299	&	207.6	&	-6.59	&	10.036	&	-9.109	&	265	&	61.7	&	4.5	&	5.5	&	79	&	0.62	&		\\
\rm Alessi\,20	&	2.970330	&	58.532158	&	423.4	&	-5.74	&	7.904	&	-2.378	&	9	&	240.4	&	11.1	&	8.7	&	349	&	0.32	&	\rm f1	\\
\rm Alessi\,20\,gp1$^a$	&	356.035534	&	56.078582	&	411.9	&	-10.68	&	8.130	&	-3.347	&	12	&	173.2	&	13.6	&	7.8	&	265	&	0.18	&	\rm f2	\\
\rm Alessi\,20\,isl1$^a$	&	11.906507	&	50.262109	&	458.9	&	-3.21	&	7.391	&	-3.108	&	100	&	97.6	&	7.5	&	6.4	&	107	&	0.41	&		\\
\rm Alessi\,24	&	260.842675	&	-62.670437	&	484.2	&	12.10	&	-0.441	&	-8.948	&	88	&	119.1	&	4.6	&	6.9	&	138	&	0.73	&		\\
\rm Alessi\,62	&	284.025507	&	21.581477	&	618.9	&	12.77	&	0.256	&	-1.120	&	692	&	144.1	&	5.7	&	7.3	&	139	&	0.63	&		\\
\rm ASCC\,16	&	81.160907	&	1.558096	&	346.4	&	20.95	&	1.299	&	-0.013	&	11	&	241.7	&	6.0	&	8.7	&	373	&	0.74	&		\\
\rm ASCC\,19	&	82.011177	&	-1.975022	&	354.7	&	24.50	&	1.118	&	-1.173	&	9	&	198.5	&	7.8	&	8.2	&	323	&	0.53	&	\rm f2	\\
\rm ASCC\,58	&	153.712881	&	-55.126490	&	477.1	&	11.69	&	-13.353	&	2.703	&	52	&	254.9	&	9.9	&	8.9	&	328	&	0.44	&	\rm f2	\\
\rm ASCC\,105	&	295.444115	&	27.493132	&	557.8	&	-15.46	&	1.426	&	-1.602	&	74	&	67.5	&	5.9	&	5.7	&	67	&	0.48	&	\rm f1	\\
\rm ASCC\,127	&	346.610023	&	65.131092	&	374.4	&	-11.89	&	7.424	&	-1.785	&	15	&	183.0	&	11.9	&	7.9	&	260	&	0.29	&	\rm f1	\\
\rm BH\,99	&	159.413798	&	-59.062415	&	446.9	&	13.08	&	-14.470	&	1.089	&	81	&	556.7	&	8.3	&	11.5	&	730	&	0.63	&	\rm f2	\\
\rm BH\,164	&	222.290304	&	-66.446488	&	419.8	&	-0.03	&	-7.404	&	-10.689	&	65	&	194.5	&	5.3	&	8.1	&	243	&	0.72	&		\\
\rm Blanco\,1$^b$	&	1.536331	&	-29.975190	&	236.4	&	6.21	&	18.708	&	2.606	&	100	&	342.9	&	6.7	&	10.2	&	703	&	0.64	&	\rm t	\\
\rm Collinder\,69	&	83.809113	&	9.809913	&	399.0	&	25.89	&	1.300	&	-2.113	&	14	&	401.8	&	8.1	&	10.3	&	620	&	0.63	&	\rm f1	\\
\rm Collinder\,135$^b$	&	109.553612	&	-37.079084	&	302.8	&	16.12	&	-10.101	&	6.165	&	40	&	253.2	&	9.7	&	8.9	&	377	&	0.41	&	\rm f2	\\
\rm Collinder\,140	&	110.944634	&	-32.063703	&	384.5	&	19.07	&	-8.073	&	4.769	&	50	&	179.6	&	10.2	&	7.9	&	241	&	0.42	&	\rm f2	\\
\rm Collinder\,350	&	267.003004	&	1.452913	&	367.8	&	-14.79	&	-4.933	&	-0.060	&	589	&	149.4	&	6.6	&	7.4	&	156	&	0.51	&	\rm t	\\
\rm Coma\,Berenices$^b$	&	186.034554	&	25.539697	&	85.6	&	0.04	&	-12.018	&	-8.940	&	700	&	101.6	&	4.7	&	6.8	&	158	&	0.59	&	\rm t	\\
\rm Group\,X	&	217.766195	&	54.432658	&	99.6	&	-6.50	&	-16.051	&	-2.802	&	400	&	99.5	&	14.5	&	6.8	&	187	&	0.11	&	d	\\
\rm Gulliver\,6	&	83.207952	&	-1.711134	&	413.3	&	30.27	&	0.128	&	-0.269	&	7	&	168.5	&	7.3	&	7.7	&	284	&	0.54	&	\rm f1	\\
\rm Gulliver\,21	&	106.889853	&	-25.471498	&	652.5	&	41.25	&	-1.875	&	4.226	&	275	&	83.3	&	4.9	&	6.1	&	74	&	0.68	&		\\
\rm Huluwa\,1$^b$	&	122.390262	&	-47.017897	&	354.7	&	17.00	&	-6.374	&	9.360	&	12	&	724.0	&	14.6	&	12.6	&	1294	&	0.39	&	\rm f1	\\
\rm Huluwa\,2$^b$	&	121.163632	&	-48.884799	&	398.8	&	20.65	&	-5.525	&	8.226	&	11	&	467.4	&	15.3	&	10.9	&	743	&	0.26	&	\rm f1	\\
\rm Huluwa\,3$^b$	&	117.397635	&	-46.600642	&	394.7	&	21.76	&	-4.706	&	8.960	&	11	&	372.8	&	7.9	&	10.1	&	588	&	0.66	&		\\
\rm Huluwa\,4$^b$	&	125.994565	&	-41.178685	&	341.8	&	19.83	&	-7.115	&	10.021	&	10	&	180.7	&	14.8	&	7.9	&	347	&	0.28	&	\rm f1	\\
\rm Huluwa\,5$^b$	&	126.933768	&	-34.930350	&	355.0	&	15.69	&	-7.007	&	10.779	&	8	&	60.8	&	4.7	&	5.5	&	102	&	0.65	&	\rm f1	\\
\rm IC\,348	&	56.093028	&	32.173637	&	316.5	&	16.99	&	4.415	&	-6.407	&	5	&	142.2	&	4.1	&	7.3	&	211	&	0.73	&		\\
\rm IC\,2391$^b$	&	130.229784	&	-52.990392	&	151.3	&	15.01	&	-24.641	&	23.309	&	50	&	140.2	&	2.5	&	7.6	&	219	&	0.99	&		\\
\rm IC\,2602$^b$	&	160.515429	&	-64.443674	&	151.4	&	16.57	&	-17.691	&	10.695	&	45	&	188.1	&	3.7	&	8.4	&	318	&	0.91	&		\\
\rm IC\,4665$^b$	&	266.568565	&	5.608075	&	347.4	&	-13.19	&	-0.845	&	-8.543	&	36	&	158.5	&	6.0	&	7.9	&	197	&	0.72	&		\\
\rm IC\,4756	&	279.617685	&	5.430164	&	473.7	&	-24.37	&	1.236	&	-4.991	&	955	&	508.0	&	8.2	&	11.2	&	497	&	0.70	&	\rm t	\\
\rm LP\,2371	&	82.056845	&	1.632299	&	367.0	&	29.36	&	-0.549	&	0.746	&	20	&	81.0	&	5.1	&	6.1	&	103	&	0.67	&		\\
\rm LP\,2373	&	83.844838	&	-5.670678	&	386.9	&	21.01	&	1.621	&	-1.097	&	9	&	98.0	&	7.9	&	6.5	&	153	&	0.39	&	\rm f1	\\
\rm LP\,2373\,gp1$^a$	&	81.170979	&	-2.515377	&	335.6	&	22.52	&	1.104	&	-0.277	&	11	&	187.4	&	12.0	&	8.0	&	341	&	0.23	&	\rm f1	\\
\rm LP\,2373\,gp2	&	81.836831	&	1.892265	&	349.3	&	17.77	&	1.543	&	-0.545	&	9	&	543.0	&	11.6	&	11.4	&	884	&	0.47	&	\rm f1	\\
\rm LP\,2373\,gp3	&	83.509434	&	-0.498267	&	349.1	&	20.10	&	1.671	&	-0.932	&	6	&	111.3	&	9.0	&	6.7	&	177	&	0.34	&	\rm f1	\\
\rm LP\,2373\,gp4	&	84.260923	&	-2.053777	&	363.0	&	20.67	&	1.708	&	-1.310	&	6	&	296.2	&	9.9	&	9.3	&	490	&	0.48	&	\rm f2	\\
\rm LP\,2383	&	95.335972	&	-16.218225	&	364.3	&	24.81	&	-5.420	&	5.068	&	50	&	280.7	&	12.8	&	9.2	&	475	&	0.34	&	\rm f2	\\
\rm LP\,2388	&	127.066134	&	-47.931643	&	497.3	&	24.94	&	-5.923	&	6.884	&	200	&	149.4	&	6.8	&	7.4	&	199	&	0.53	&		\\
\rm LP\,2428	&	43.296942	&	68.875038	&	436.1	&	-1.96	&	1.428	&	-7.976	&	200	&	111.0	&	6.0	&	6.7	&	142	&	0.58	&		\\
\rm LP\,2429	&	84.460533	&	57.169259	&	479.5	&	-6.36	&	-3.222	&	-3.992	&	1150	&	148.0	&	6.1	&	7.4	&	179	&	0.61	&	\rm t	\\
\rm LP\,2439	&	103.540197	&	-5.896183	&	283.9	&	23.13	&	-7.336	&	-2.457	&	25	&	143.0	&	6.5	&	7.3	&	243	&	0.54	&	\rm f2	\\
\rm LP\,2441	&	279.297560	&	-14.225387	&	280.4	&	-21.37	&	-1.772	&	-9.437	&	75	&	187.8	&	8.9	&	8.0	&	226	&	0.46	&	\rm f2	\\
\rm LP\,2442	&	250.101032	&	-39.560185	&	175.8	&	1.45	&	-11.929	&	-21.288	&	15	&	318.7	&	6.2	&	9.6	&	662	&	0.70	&	\rm f2	\\
\rm LP\,2442\,gp1$^a$	&	242.331352	&	-22.907537	&	138.9	&	-7.47	&	-11.731	&	-24.786	&	8	&	111.6	&	8.7	&	6.7	&	220	&	0.29	&	\rm f2	\\
\rm LP\,2442\,gp2$^a$	&	243.556287	&	-23.095881	&	141.1	&	-6.73	&	-8.884	&	-24.562	&	8	&	151.8	&	7.9	&	7.5	&	363	&	0.47	&	\rm f2	\\
\rm LP\,2442\,gp3$^a$	&	241.119806	&	-22.113086	&	143.4	&	-6.21	&	-12.199	&	-23.174	&	8	&	64.0	&	6.4	&	5.6	&	136	&	0.38	&	\rm f2	\\
\rm LP\,2442\,gp4$^a$	&	244.532627	&	-25.119324	&	153.7	&	-4.34	&	-10.774	&	-21.905	&	8	&	113.0	&	10.2	&	6.8	&	284	&	0.26	&	\rm f2	\\
\rm LP\,2442\,gp5$^a$	&	242.398266	&	-20.348396	&	153.9	&	-5.61	&	-9.609	&	-21.354	&	8	&	76.9	&	9.8	&	6.0	&	163	&	0.26	&	\rm f2	\\
\rm Mamajek\,4	&	276.232342	&	-51.301191	&	449.9	&	-28.90	&	4.413	&	-21.456	&	372	&	281.5	&	17.7	&	9.2	&	320	&	0.21	&	\rm t	\\
\rm NGC\,1901	&	79.555700	&	-68.181508	&	417.8	&	0.67	&	1.681	&	12.694	&	850	&	124.8	&	6.7	&	7.0	&	144	&	0.52	&		\\
\rm NGC\,1977	&	83.847140	&	-4.973568	&	392.3	&	25.28	&	1.120	&	-0.547	&	3	&	107.7	&	4.7	&	6.7	&	202	&	0.76	&		\\
\rm NGC\,1980	&	83.804049	&	-5.865419	&	383.6	&	23.50	&	1.119	&	0.344	&	6	&	753.2	&	7.6	&	12.7	&	1298	&	0.81	&		\\
\rm NGC\,2232$^b$	&	96.721725	&	-4.630480	&	320.0	&	26.37	&	-4.626	&	-1.851	&	25	&	205.8	&	6.8	&	8.6	&	281	&	0.57	&	\rm f1	\\
\rm NGC\,2422$^b$	&	114.122352	&	-14.484943	&	476.4	&	35.86	&	-7.051	&	1.026	&	73	&	480.2	&	4.9	&	11.4	&	466	&	0.90	&		\\
\rm NGC\,2451A$^b$	&	115.754528	&	-38.264155	&	192.6	&	22.62	&	-20.954	&	15.333	&	58	&	182.2	&	4.9	&	8.3	&	311	&	0.82	&		\\
\rm NGC\,2451B	&	116.079334	&	-38.016157	&	363.2	&	16.65	&	-9.575	&	4.846	&	50	&	276.1	&	8.0	&	9.1	&	400	&	0.54	&	\rm f2	\\
\rm NGC\,2516$^b$	&	119.492636	&	-60.786856	&	410.6	&	24.22	&	-4.620	&	11.203	&	123	&	1973.3	&	7.9	&	18.3	&	2690	&	0.88	&	\rm t	\\
\rm NGC\,2547$^b$	&	122.490342	&	-49.178877	&	387.2	&	13.42	&	-8.562	&	4.365	&	40	&	303.9	&	5.6	&	9.8	&	452	&	0.71	&	\rm f2	\\
\rm NGC\,3228	&	155.377537	&	-51.880903	&	482.2	&	16.49	&	-14.848	&	-0.573	&	63	&	84.4	&	6.3	&	6.1	&	81	&	0.50	&		\\
\rm NGC\,3532	&	166.388592	&	-58.701886	&	478.3	&	5.44	&	-10.400	&	5.235	&	398	&	2210.5	&	7.4	&	18.2	&	2559	&	0.90	&	\rm h	\\
\rm NGC\,6405	&	265.095354	&	-32.227742	&	457.6	&	-8.12	&	-1.383	&	-5.859	&	79	&	596.9	&	5.3	&	11.8	&	654	&	0.87	&		\\
\rm NGC\,6475	&	268.466733	&	-34.828885	&	279.5	&	-14.55	&	3.083	&	-5.403	&	186	&	1024.5	&	6.1	&	14.1	&	1157	&	0.86	&	\rm h	\\
\rm NGC\,6633$^b$	&	276.890553	&	6.680996	&	394.0	&	-28.11	&	1.242	&	-1.800	&	426	&	337.3	&	5.3	&	10.2	&	300	&	0.82	&		\\
\rm NGC\,6774$^b$	&	289.091567	&	-16.317354	&	306.4	&	42.24	&	-0.889	&	-26.656	&	2650	&	152.6	&	4.7	&	7.8	&	154	&	0.76	&		\\
\rm NGC\,7058	&	320.451323	&	50.821327	&	365.2	&	-19.21	&	7.483	&	2.772	&	80	&	126.8	&	3.4	&	7.0	&	190	&	0.83	&		\\
\rm NGC\,7092	&	322.782191	&	48.247177	&	297.1	&	-5.29	&	-7.387	&	-19.718	&	350	&	193.1	&	6.5	&	8.1	&	303	&	0.61	&	\rm t	\\
\rm Pleiades$^b$	&	56.637062	&	24.132371	&	135.9	&	5.72	&	19.945	&	-45.365	&	125	&	744.0	&	4.7	&	12.7	&	1407	&	0.86	&	\rm t	\\
\rm Praesepe	&	130.008023	&	19.601297	&	185.0	&	35.00	&	-35.904	&	-12.864	&	700	&	601.8	&	4.7	&	11.8	&	982	&	0.92	&	\rm h	\\
\rm Roslund\,5	&	302.659299	&	33.751082	&	540.5	&	-16.61	&	2.078	&	-1.183	&	98	&	191.5	&	6.8	&	8.1	&	207	&	0.57	&	\rm f2	\\
\rm RSG\,7	&	343.949646	&	59.772583	&	419.3	&	-8.55	&	4.760	&	-2.244	&	70	&	67.6	&	4.6	&	5.7	&	69	&	0.59	&		\\
\rm RSG\,8	&	345.689844	&	59.113059	&	474.6	&	-8.68	&	5.665	&	-1.663	&	18	&	342.8	&	22.0	&	9.8	&	408	&	0.09	&	\rm f2	\\
\rm Stephenson\,1	&	283.406739	&	36.720713	&	358.2	&	-19.13	&	1.075	&	-3.020	&	46	&	263.4	&	7.0	&	9.0	&	407	&	0.58	&	\rm f1	\\
\rm Stock\,1	&	294.146449	&	25.145437	&	406.3	&	-19.75	&	6.105	&	0.205	&	470	&	136.3	&	4.6	&	7.2	&	125	&	0.77	&		\\
\rm Stock\,12	&	353.914758	&	52.678590	&	437.7	&	-2.18	&	8.553	&	-1.920	&	112	&	122.0	&	4.3	&	6.9	&	124	&	0.73	&		\\
\rm Stock\,23	&	49.134918	&	60.355183	&	606.7	&	-18.57	&	-4.313	&	-1.010	&	94	&	106.0	&	9.1	&	6.6	&	78	&	0.37	&	\rm f1	\\
\rm UBC\,7$^b$	&	106.935171	&	-37.677240	&	278.3	&	16.78	&	-9.730	&	7.004	&	40	&	192.3	&	7.6	&	8.1	&	336	&	0.54	&	\rm f2	\\
\rm UBC\,19	&	56.443686	&	29.936422	&	399.7	&	19.45	&	2.825	&	-5.217	&	7	&	42.2	&	6.1	&	4.9	&	57	&	0.38	&		\\
\rm UBC\,31	&	60.707033	&	32.542298	&	365.0	&	20.89	&	3.698	&	-5.370	&	12	&	260.0	&	14.7	&	8.9	&	353	&	0.13	&	\rm f1	\\
\rm UBC\,31\,gp1$^a$	&	63.563584	&	32.027203	&	339.2	&	20.90	&	3.766	&	-5.937	&	12	&	58.9	&	10.7	&	5.4	&	80	&	0.13	&	\rm f1	\\
\rm UBC\,31\,gp2$^a$	&	57.959044	&	35.028000	&	381.4	&	23.12	&	3.385	&	-4.450	&	10	&	185.1	&	10.7	&	8.0	&	213	&	0.27	&	\rm f1	\\
\rm UPK\,82	&	298.265305	&	26.504003	&	542.2	&	-12.72	&	2.258	&	-2.117	&	81	&	57.6	&	4.8	&	5.4	&	61	&	0.56	&		\\

	\enddata
	\tablecomments{\\
	    $^a$ New hierarchical groups identified in this work. \\
	    $^b$ Clusters members taken from \citet{pang2021a,pang2021b,li2021}.\\ Columns 2--7 list the median value of cluster members. R.A. and Decl. are the right ascension and declination. Dist. is the distance after correction in Section~\ref{sec:dis_corr}. RV is the radial velocity. \pmra and \pmdec  \,\,are the components of the proper motion. The age of the cluster is derived from PARSEC isochrone fitting. $M_{\rm cl}$ is the total mass of each star cluster. $r_{\rm h}$ and $r_{\rm t}$ are the half-mass and the tidal radii of each cluster, respectively. $N$ is the total number of members, and $f_m$ is the bound mass fraction, which is the ratio of mass within the tidal radius. The flag in column~14 indicates the morphological type of the cluster substructures outside $r_{\rm t}$: ``f1'' for filamentary, ``f2'' for fractal, ``h'' for halo, and ``t'' for  tidal-tail (see Section~\ref{sec:class} for a further description). Group\,X is instead labeled as ``d'' for its total disruption state \citep{tang2019}.
	}
\end{deluxetable*}

\begin{deluxetable*}{ccc}
\tablecaption{Columns for the table of individual members of the 65 clusters identified in this work. \label{tab:memberlist}
             }
\tabletypesize{\scriptsize}
\tablehead{
\colhead{Column}    & \colhead{Unit}    & \colhead{Description}
}
\startdata
Cluster Name                    & -                 &  Name of the target cluster   \\
      Gaia ID                   & -                 &  Object ID in Gaia EDR\,3\\
ra                              & degree           &  R.A. at J2016.0 from Gaia EDR\,3\\
er\_RA                          & mas              &  Positional uncertainty in R.A. at J2016.0 from Gaia EDR\,3 \\
dec                             & degree           &  Decl. at J2016.0 from Gaia EDR\,3 \\
er\_DEC                         & mas              &  Positional uncertainty in decl. at J2016.0 from Gaia EDR\,3 \\
parallax                        & mas              &  Parallax from Gaia EDR\,3\\
er\_parallax                    & mas              &  Uncertainty in the parallax \\
pmra                            & mas~yr$^{-1}$    &  Proper motion with robust fit in $\alpha \cos\delta$ from {\it Gaia} EDR\,3     \\
er\_pmra                        & mas~yr$^{-1}$    &  Error of the proper motion with robust fit in $\alpha \cos\delta$   \\
pmdec                           & mas~yr$^{-1}$    &  Proper motion with robust fit in $\delta$ from Gaia EDR\,3     \\
er\_pmdec                       & mas~yr$^{-1}$    &  Error of the proper motion with robust fit in $\delta$  \\
Gmag                            & mag              & Magnitude in $G$ band from Gaia EDR\,3   \\
BR                              & mag              & Magnitude in $BR$ band from Gaia EDR\,3   \\
RP                              & mag              & Magnitude in $RP$ band from Gaia EDR\,3   \\
Gaia\_radial\_velocity          & km~s$^{-1}$      &  Radial velocity from Gaia DR\,2 \\
er\_Gaia\_radial\_velocity      & km~s$^{-1}$      &  Error of radial velocity from Gaia DR\,2\\
Mass                            & M$_\odot$        & Stellar mass obtained in this study\\
X\_obs                          & pc               & Heliocentric Cartesian X coordinate computed via direct inverting Gaia EDR\,3 parallax $\varpi$ \\
Y\_obs                          & pc               & Heliocentric Cartesian Y coordinate computed via direct inverting Gaia EDR\,3 parallax $\varpi$ \\
Z\_obs                          & pc               & Heliocentric Cartesian Z coordinate computed via direct inverting Gaia EDR\,3 parallax $\varpi$ \\
X\_cor                          & pc               & Heliocentric Cartesian X coordinate after distance correction in this study \\
Y\_cor                          & pc               & Heliocentric Cartesian Y coordinate after distance correction in this study \\
Z\_cor                          & pc               & Heliocentric Cartesian Z coordinate after distance correction in this study \\
Dist\_cor                       & pc               & The corrected distance of individual member\\
\enddata
\tablecomments{A machine-readable version of this table is available online.
    }
\end{deluxetable*}

\subsection{Membership Verification}\label{sec:member_reliablity}

Accurate cluster membership is crucial for interpreting the morphology of an open cluster. Before further analysis, we verify the reliability of member star list obtained in Section~\ref{sec:stargo} using $N$-body numerical models to produce simulated cluster members. We exempt Group\,X from $N$-body simulations, since it requires additional treatment to reproduce its two-component fragmented structures in space. The simulated clusters produced in this section are used to verify membership and later on to distance correction in Section~\ref{sec:dis_corr}. 

\subsubsection{$N$-body simulations}\label{sec:nbody}

We use the high-performance $N$-body cluster simulation code \texttt{petar} \citep{Wang2020b}, accelerated by \texttt{fdps} \citep{iwasawa2015,iwasawa2016,iwasawa2017}, to simulate the evolution of clusters. The \texttt{sdar} code \citep{wang2020a} is embedded in \texttt{petar} to ensure that the close encounters and the orbits of binaries are accurately treated in the simulation.
All cluster models are initialized with a Plummer profile in virial equilibrium, characterized by the initial cluster mass $M_{\rm cl}(0)$ and the initial half-mass radius $r_{\rm h,i}$. Stellar masses are sampled from the \citet{kroupa2001} initial mass function (IMF) in the mass range  $0.08-150\,M_\odot$ with optimal sampling applied \citep{kroupa2013, yan2017}. The total mass of a simulated cluster, $M_{\rm cl}(0)$, is estimated from the present-day mass and adding the cumulative wind mass loss of all observed stars using the \texttt{SSE} code \citep{Hurley2000,banerjee2020}. 
For each observed cluster, we only perform one simulation with estimations for the initial conditions (it is not the purpose of this study to exactly replicate the observed clusters). \cite{Wang2021a} demonstrated that the large statistical variations in the initial number of OB stars due to the random sampling from the IMF can significantly affect the long-term evolution of the cluster. Thus, the optimal sampling from the IMF is suitable to reduce the stochastic effect. In addition, \cite{Wang2021a} shows that the initial half-mass radius, $r_{\rm h,i}$, does not influence the evolution much, thus for the current purpose we simply assume $r_{\rm h}$ to be 1\,pc at the start of the simulations.

We date back to the location where each cluster was born using the Python package \texttt{Galpy} \citep{Bovy2015} with the knowledge about its current positions and 3D velocity (see Table~\ref{tab:paramter_all}). The adopted simulation model starts after the gas expulsion phase has ended, and contains only stellar components. Although \texttt{petar} can accurately handle the interactions between binaries and single stars \citep{wang2020a}, we do not include primordial binary systems, as they do not affect the main purpose of this simulation\,---\,to verify the accuracy of membership of cluster members. The incompleteness originating from binary stars is beyond the scope of the current study.

The \texttt{petar} code adopted algorithms from \citet{Jerabkova2021}, \citet{Wang2021a} and \texttt{PyGaia} package\footnote{Developed by Tommaso Marchetti \url{https://github.com/agabrown/PyGaia.}} to synthesize observational Gaia photometry and parallaxes. The parallax errors are added to the simulation results based on the observational data to regenerate the artificial line-of-sight elongation. The outputs of \texttt{petar} include 3D spatial positions: coordinates and parallax; 3D motions: PMs and RV; and Gaia $G$-band magnitude.

\subsubsection{Verification} 

Among the simulated clusters, we select three representative clusters, Pleiades, NGC\,3532, and Blanco\,1, which have distinguishing  morphological features, to verify the robustness of our membership identification method. The Pleiades is an intermediate-mass cluster with only early tidal disruption signature observed \citep{li2021}. NGC\,3532, on the other hand, is the most massive cluster in our sample, and has a very dense core. Blanco\,1 is unique for having the lowest cluster mass among these three clusters ($\sim$15\% of the mass of NGC\,3532) but well known for its two 50--60\,pc long tidal tails \citep{Zhang2019}.

To compare with observations, we select members located within 100\,pc from the cluster center for further verification. From the PM distribution diagram \citepalias[see Figure~1 in][]{pang2021a}, we define a rectangular region that confines the 100\,pc members. With the simulated cluster members ready, we add the Gaia EDR\,3 mock field stars from \citet{rybizki2020} to form a simulated observation data set.
The mock field stars are selected from the pre-defined rectangular region on the PM space by the 100\,pc members. With the knowledge of real simulated cluster members and real field stars in this simulated observational data set, we are now able to assess the reliability of cluster membership identified from the process described in Section~\ref{sec:stargo}.


\begin{figure*}[!tbh]
\centering
\includegraphics[angle=0, width=1.0\textwidth]{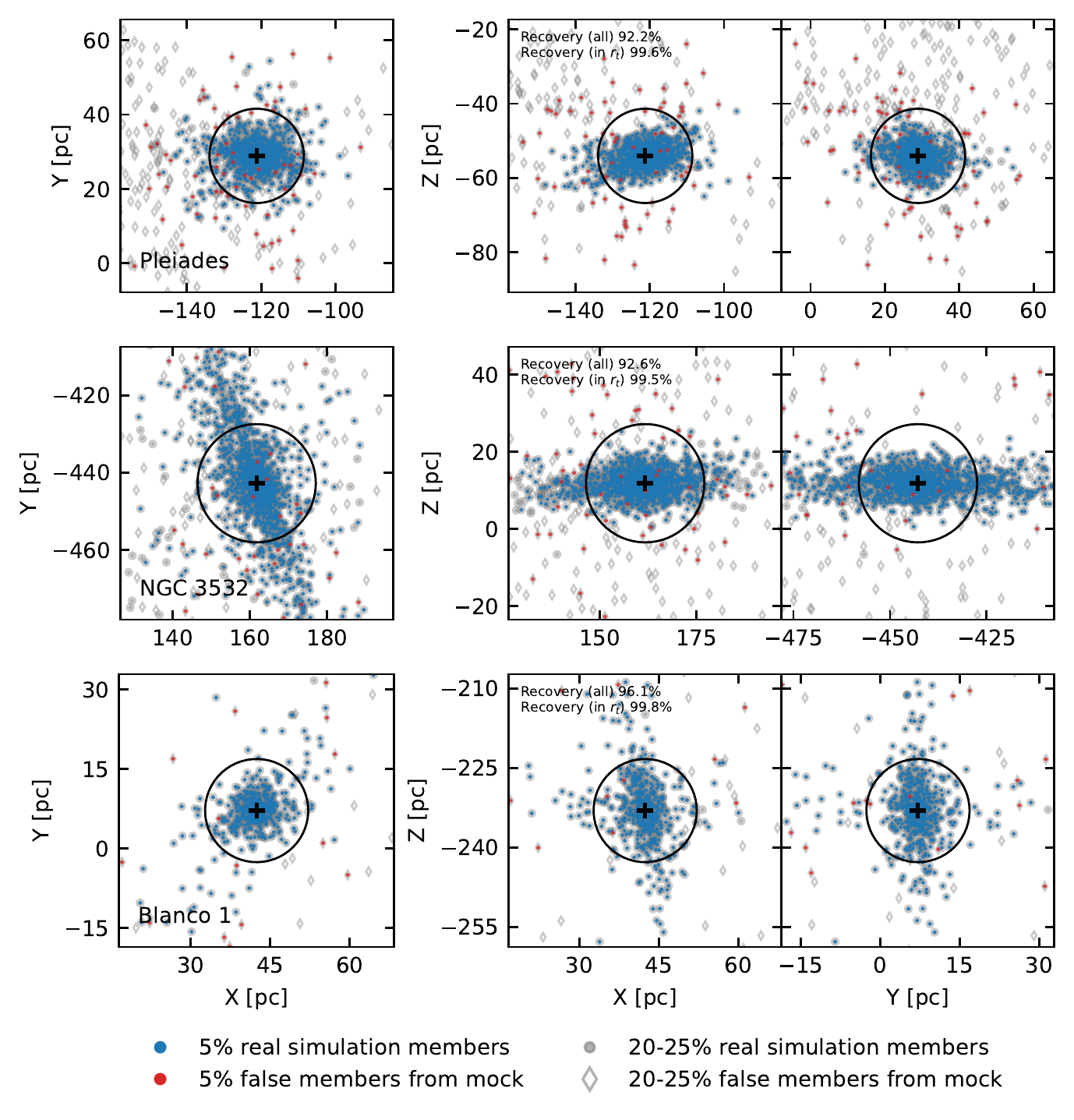}
    \caption{
        Spatial distribution of simulated cluster members identified by \texttt{StarGo} for three example clusters: Pleiades (upper panels), NGC\,3532 (middle panels), and Blanco\,1 (bottom panels). The blue dots are real members of simulated clusters and recovered by \texttt{StarGo} at a 5\% contamination rate. The red dots are contaminant mock field stars falsely identified as members (false positives). The background grey dots and diamonds are real and false members obtained with 20--25\% contamination rate, respectively. The black circle represents the tidal radius of each cluster.  
        }
\label{fig:nbody}
\end{figure*}

In Figure~\ref{fig:nbody} we show the 3D spatial distribution of the simulated cluster members recovered by \texttt{StarGo} with a 5\% percent contamination rate (blue dots) of the three representative clusters. Mock field stars that are falsely identified as members (false positives) indicated in red. The recovery rates of simulated members are above 92 percent in all three  clusters. Because Blanco\,1 is located below the Galactic plane where field star density is lower, it has the highest recovery rate $\sim$96\%. Elongated substructures, such as tidal tails, are well recovered for Blanco\,1. When the contamination rate rises up to 20--25\%, the number of false cluster members (grey diamonds) climbs accordingly. These false-positive stars are mainly positioned randomly in the cluster's outer region, forming an artificial ``halo''-like structure around the cluster. This fake ``halo'' feature becomes more prominent in sky regions having a higher field star density, e.g., NGC\,3532. One should be cautious with interpreting the overall cluster morphology if the member contamination rate is high ($\gtrsim$20\%).

In the contrast, member recovery rate inside the tidal radius \citep[$r_{\rm t}$, computed using equation~12 in][]{pinfield1998}, is quite stable and not affected by the overall contamination rate. The bound members (stars within $r_{\rm t}$) are successfully recovered by more than 99.5\% in all three example clusters. Even when the overall contamination rate raised to 20--25\%, the simulated cluster members identified within the $r_{\rm t}$ still overwhelm the false positives, with a recovery rate of 99\%. Therefore, this high recovery rate within the $r_{\rm t}$ shows the robustness of our cluster membership identification within $r_{\rm t}$ under a contamination rate of 25\%. 

In the following sections, we will derive the overall morphology of clusters via the 5\% contamination rate members in a qualitative manner (Section~\ref{sec:3D_solar}). To ensure an unbiased analysis of the 3D shapes of all clusters, we only use members within $r_{\rm t}$ for detailed morphological quantification (ellipsoid fitting in Section~\ref{sec:mor_tidal}).

\begin{figure*}[tbh]
\centering
\includegraphics[angle=0, width=0.95\textwidth]{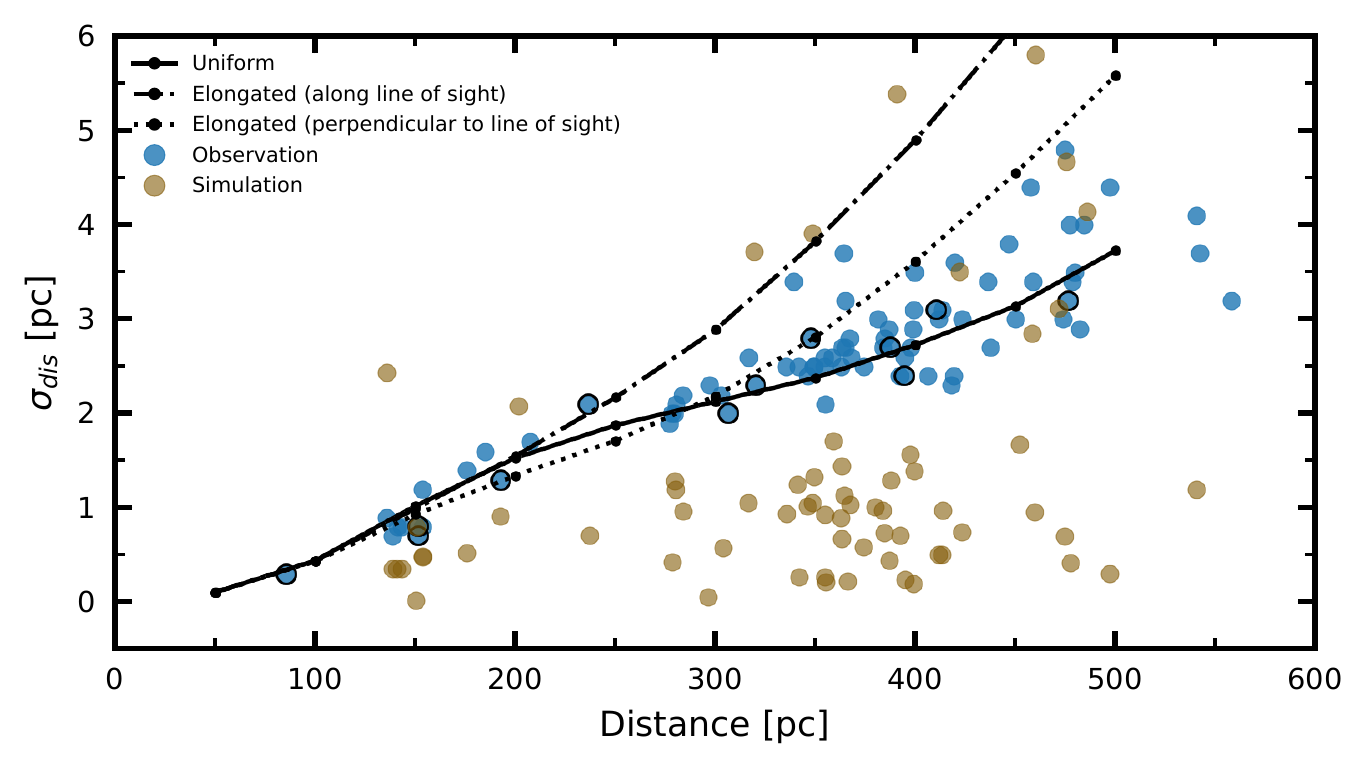}
\caption{
    Dependence of the uncertainty in the corrected distance $\sigma_{\rm dis}$ on cluster distances. The black solid curve represents a star cluster in which the members have a uniform spatial distribution (See Appendix~C in \citetalias{pang2021a}). The dotted and dashed-dotted curves are clusters with elongated shape perpendicular to and align with the line-of-sight, respectively. The blue dots indicate errors in the corrected distances when adopting a star cluster with a uniform spherical stellar density located at the real distance of each cluster in our study, except Group\,X. Black open circles indicate cluster with their $\sigma_{\rm dis}$ value taken from \citetalias{pang2021a}. The olive points are simulated clusters having more than 10 member stars inside the tidal radius, to which distance correction is applied. The value of $\sigma_{\rm dis}$ is generally lower for simulated clusters owing to a shallower density profile than the uniform spatial distribution.
    }
\label{fig:dis-error}
\end{figure*}

\section{Distance correction}\label{sec:dis_corr}

An apparent stretching along the line-of-sight in open clusters results from obtaining stellar distances from inverting the parallax; this phenomenon has been well studied \citep[e.g.,][]{carrera2019,Zhang2019,pang2020,pang2021a,tarricq2021,piecka2021}.
\citet{pang2020,pang2021a} attempted to mitigate this problem via a Bayesian approach described in \citet{carrera2019}. This Bayesian approach is based on the assumption of a normal distribution for the individual distances to the cluster stars and an exponentially decreasing profile for the distances to field stars \citep{bailer2015}. To understand the limitation of this correction method, \citetalias{pang2021a} carried out a Monte-Carlo (MC) simulation and showed that the uncertainty in the corrected distances, $\sigma_{\rm dis}$, increases monotonically with distance\footnote{The $\sigma_{\rm dis}$ estimated for each cluster is obtained using a spherical uniform stellar density profile and placed at the real distance of each cluster (further details in Appendix~C of \citetalias{pang2021a}).}. In the MC simulations, the $\sigma_{\rm dis}$ reaches an accepted value of 3.0--6.3\,pc at a distance of 500\,pc under two initial conditions: (1) a spherical uniform distribution of cluster members, and (2) a cluster with members artificially elongated (perpendicular or along with line-of-sight). We adopt the same Bayesian procedure to correct the distances of all the members of the 65 clusters in this work. The corrected distances and the 3D positions are listed in Table~\ref{tab:memberlist}, together with other parameters from Gaia EDR\,3. Group\,X is excluded from distance correction because of its fragmented spatial distribution, and, thus, the spatial positions of its members are directly computed from inverse Gaia parallax.

In Figure~\ref{fig:dis-error}, we present the $\sigma_{\rm dis}$ (blue dots) for all 84 OCs (twelve taken from \citetalias{pang2021a} are highlighted with black open circles). The value of $\sigma_{\rm dis}$ resides mostly between the predicted value of uniform and elongated models (solid and dashed-dotted curves), depending on the quality of the parallax. Clusters at a further distance suffer from a reduced parallax quality. However, the value of the tidal radius of each cluster is typically three times larger than the uncertainty in the corrected distance $\sigma_{\rm dis}$, ensuring an accurate determination of the intrinsic cluster morphology within the tidal radius.

We apply the distance correction to the 84 simulated clusters in order to validate the reliability of the distance correction. Within the 84 simulated clusters, eighteen develop into having tidal tails that stretch out beyond a few 100\,pc. Fewer than ten stars remain in the bound region (i.e., within tidal tails) in those eighteen clusters, a sign of ongoing dispersal. Therefore, we only correct for clusters with more than ten members within $r_{\rm t}$. The difference between the corrected position and the theoretical position from \texttt{petar} is defined as the $\sigma_{\rm dis}$. Together with $\sigma_{\rm dis}$ from observation data, we present the $\sigma_{\rm dis}$ of simulated clusters (olive dots) in Figure~\ref{fig:dis-error}. The $\sigma_{\rm dis}$ for simulated clusters is, in general, smaller than that of the observations, with a few outliers distributed in the region occupied by the observations. This result assures us that the distance correction is appropriate. 

\section{3D morphology in the solar neighborhood}\label{sec:3D_solar}

\begin{figure*}[!htb]
\centering
\includegraphics[angle=0, width=0.8\textwidth]{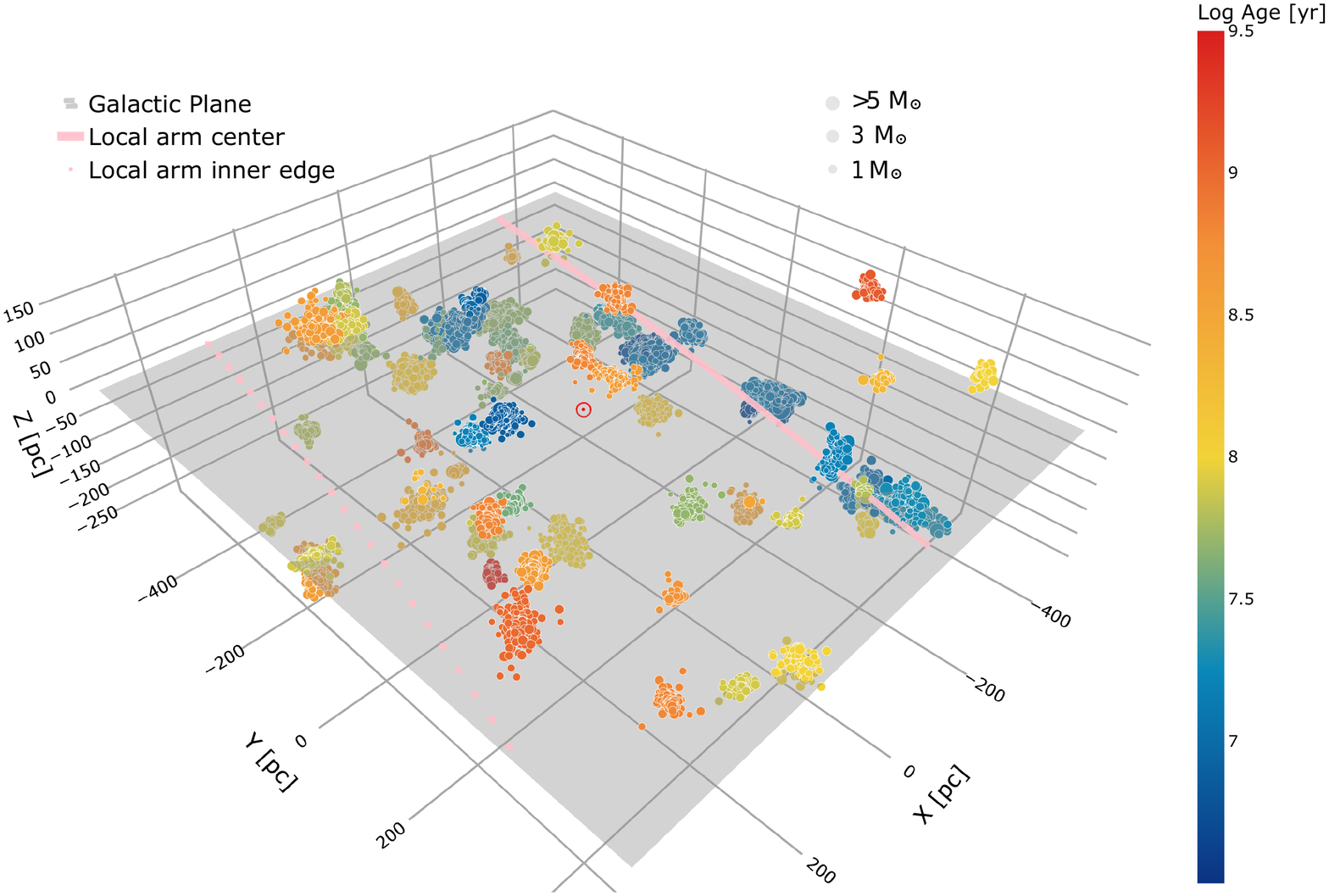}
\caption{
    3D morphology of 85 open clusters in the solar neighborhood in Heliocentric Cartesian coordinates. Each color point represents a star with its size proportional to its mass. The color of the cluster is scaled with the logarithm of age. The pink solid curve indicates the location of Local Arm center, while the pink dashed curve represent the inner edge of the Local Arm toward the Galactocentric direction \citep{reid2019}. The outer boundary of the Local Arm is outside the scope of this figure. The grey plane shows the disk of Milky Way. An interactive version of this figure is available at \url{http://3doc-morphology.lowell.edu}. It includes features (not shown here) that allow users to select individual target cluster for demonstration (by clicking on the tab ``All Clusters Overview''), or to compare two clusters (by clicking on the tab ``Compare Two Clusters'').
    }
\label{fig:xyz_3D}
\end{figure*}

In Figure~\ref{fig:xyz_3D}, we present the 3D spatial distributions for member stars in the 85 target clusters after distance correction. Our cluster samples have a larger variety in cluster total masses. The difference can be as high as two orders of magnitudes, from the lowest-mass cluster, UBC\,19,  $\sim$42\,$\rm M_\odot$ (57 members), to the most massive cluster, NGC\,3532, $\sim$2210\,$\rm M_\odot$ (2559 members). Each dot in Figure~\ref{fig:xyz_3D} represents a star, with the dot size proportional to the stellar mass (ranges from 0.1 to 13.7\,$\rm M_\odot$). An online interactive 3D demonstration is available for Figure~\ref{fig:xyz_3D}\,\footnote{ \url{http://3doc-morphology.lowell.edu}}, through which stellar mass distribution for each cluster can be observed. The age of each cluster in Figure~\ref{fig:xyz_3D} is scaled by the color bar, with a blue color representing the youngest populations in our sample and red for the oldest. Most clusters (groups) in our sample are a few to a hundred million years old. These young clusters are closely associated with the center of the Local Arm (solid pink curve), where massive star formation takes place \citep{xu2013,xu2016}. On the contrary, older generations, are approaching the edge of the Local Arm (pink dashed curve).

\begin{figure*}[!htb]
\centering
\includegraphics[angle=0, width=1.\textwidth]{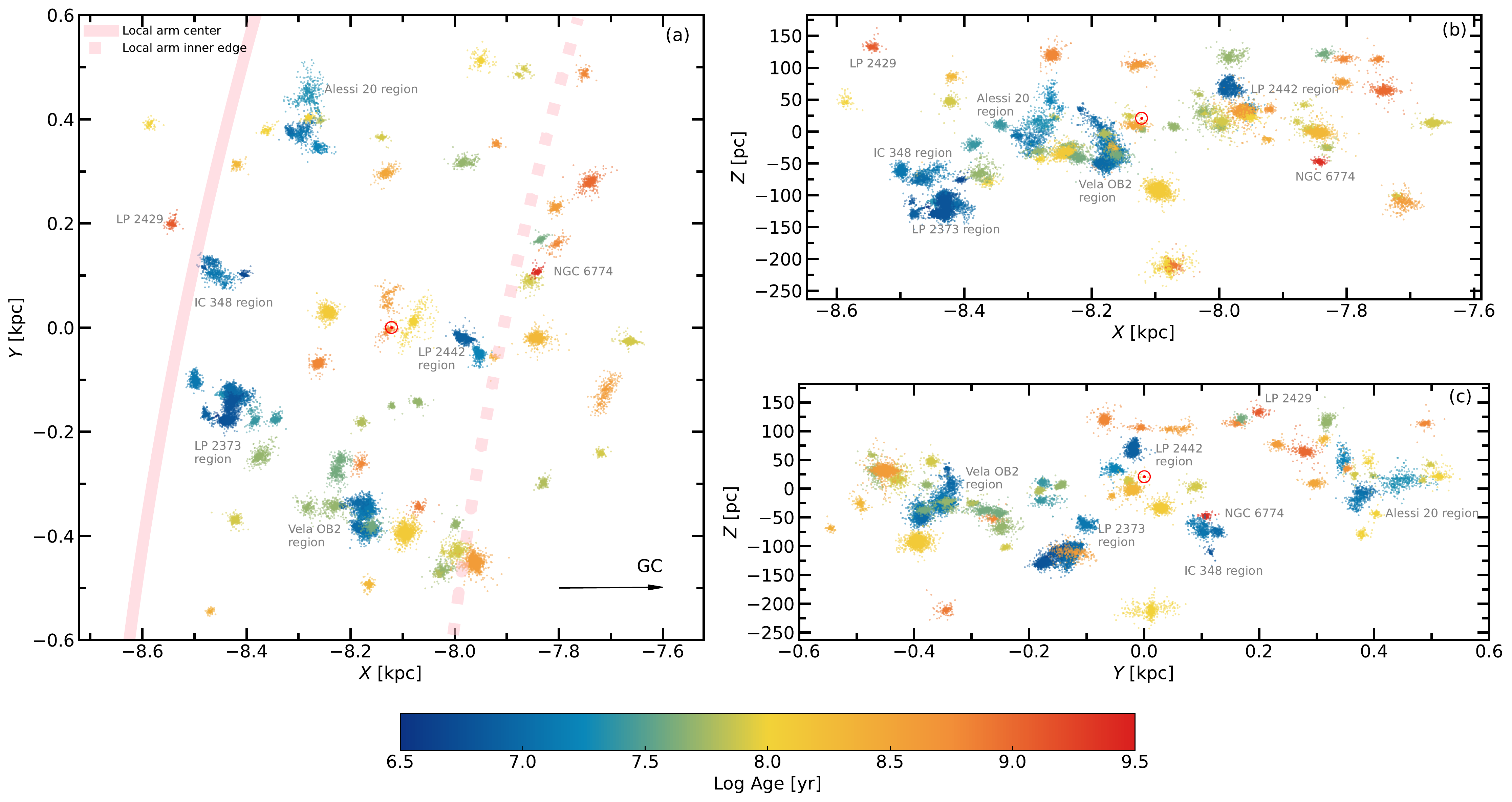}
\caption{
    3D spatial positions of member stars in all 85 clusters in Galactocentric Cartesian coordinates. Colors and symbols are identical to Figure~\ref{fig:xyz_3D}. The location of the Sun is at ($X,Y,Z$) = ($-8122$, 0, 20.8)\,pc \citep{reid2019}. Five hierarchical clustering regions and two oldest clusters are highlighted. 
    }
\label{fig:xyz_2D}
\end{figure*}

To further inspect the detailed spatial distribution of the target clusters, we project them onto $X$-$Y$, $X$-$Z$, and $Y$-$Z$ planes in Figure~\ref{fig:xyz_2D}, where we indicate   the young cluster (group) regions and typical old clusters. The clustering effect of young clusters around the center of the Local Arm (solid pink curve in Figure~\ref{fig:xyz_2D} (a)) is better observed. Younger clusters, especially those less than 30\,Myr (blue), are distributed not only closer to the center of the Local Arm but also closer to the Galactic Plane (Figure~\ref{fig:xyz_2D} (b)). On the other hand, older clusters (yellow to red, $>$100\,Myr), such as the two oldest NGC\,6774 (2.65\,Gyr) and LP\,2429 ($\sim$ 1.15\,Gyr determined in this study), show a preference for being further away from the arm and disk. The spatial positions of young clusters on the $Y$-$Z$ plane (Figure~\ref{fig:xyz_2D} (c)) appear as a wave pattern. This wave pattern resembles to some extent the ``Radcliffe'' wave found by \citet{alves2020} along the $Y$-axis, which is only associated with young clusters \citep{kounkel2020}.
The distribution of young clusters ($<$100\,Myr) in our samples on the $Y$-$Z$ plane significantly differs from a uniform distribution (a p-value of 0.02 from KS test), implying the observed wave pattern is not a random event. Although our cluster sample in the solar neighborhood is a tip of the iceberg, the spatial preference of young and old clusters is consistent with that of full-sky cluster samples identified in earlier studies \citep{soubiran_open_2018,Cantat-Gaudin2020,castro2020,hao2022}.

\subsection{Morphological Substructure Classification}\label{sec:class}

In this work, we consider the extended features outside the tidal radius of each cluster as their morphological substructure. We qualitatively classify them into four types:

\begin{enumerate}
    \item \textbf{Filamentary (f1-type)}: young clusters (groups) $<$100\,Myr with uni-directional filaments that is mostly elongated along one direction.
    \item \textbf{Fractal (f2-type)}: young clusters (groups) $<$100\,Myr with  multi-directional fractal substructures.
    \item \textbf{Halo (h-type)}: clusters with age $>$100\,Myr having a compact core but show some low-density stars spread out in the outskirt.
    \item \textbf{Tidal-tail (t-type)}: clusters older than 100\,Myr show uni-directional tidal tails.
\end{enumerate}
The classification results are listed in column~14 of Table~\ref{tab:paramter_all} and are discussed below. We label Group~X as ``d'' in Table~\ref{tab:paramter_all} to represent its disrupted state. Note that the classification does not apply to clusters (groups) without morphological features outside their tidal radii.

\subsubsection{Filamentary and Fractal Types}

Among the 60 open clusters in our sample that are younger than 100\,Myr, 41 host extended substructures outside their tidal radii. These uni-directional elongated substructures are often referred to as ``filamentary'' \citep{jerabkova2019} or ``string''-like structures \citep{kounkel2019} in literature. These types of structures are thought to be the relics of the star formation process that took place along the filaments in their parent molecular clouds (see Section~\ref{sec:intro}). Coeval relic filaments have been found in the Orion cluster \citep{jerabkova2019}, the Vela~OB2 association \citep{cantat2019a,beccari2020,pang2021b,wang2021}, and other young ($<$100\,Myr) clusters \citep{kounkel2019}.

We classify young clusters with a uni-directional filament-like substructure as the ``filamentary'' type (f1-type). Other young clusters show multi-directional fractal substructures are categorized as ``fractal'' type (f2-type). Examples of these two morphological types are shown in Figure~\ref{fig:xz_example} (in blue and red, respectively).

According to the hierarchical formation scenario, if the filamentary substructures are formed in a low star formation efficiency (SFE) region, they will disperse after gas expulsion \citep{kruijssen2012}. However, if the filamentary substructures are formed in a high-SFE region, the filaments may have a chance to converge into the central hub and then merge with other filaments to form a dense cluster \citep{vazquez-semadeni_hierarchical_2016,trevino-morales2019,ward_not_2020}. During this merger process, filaments interact and develop a fractal spatial configuration that might reflect an underlying dynamical interaction between sub-groups of stars \citep{clarke2010,arnold2017,fujii2021}. In Figure~\ref{fig:xz_example}, there is an obvious overdensity outside the central core of BH\,99 at ($X_c$, $Z_c$)$\sim$(100, 5)\, pc, resembling a sub-group in the merging process \citep{arnold2017,vazquez-semadeni_hierarchical_2016}.

\subsubsection{Halo Type}

Some clusters older than 100\,Myr that still host a dense central core show a halo of low-density stars spread out in the outskirt. We considered this substructure as ``halo'' (h-type). Only three clusters are classified as h-type, and all are shown in Figure~\ref{fig:xz_example} (purple-dot panels). Massive clusters are less susceptible to external disruption forces like the Galactic tide \citep{li2021}; thus, as these h-type clusters all have a total mass of $>$600\,$\rm M_\odot$, their current internal kinematics is probably still dominated by the two-body relaxation. This relaxation process tends to result in low-mass stars to migrate toward the outskirts of the cluster. A previous study had found tidal tails $\sim$ 200~pc long in both directions for the Praesepe cluster \citep{roser2019b}; however, because of the 100-pc limited size in the spatial cut, we can only recover its closest unbound halo substructures in this study. Praesepe is therefore classified as a halo type.

\subsubsection{Tidal-tail Type}

\begin{figure*}[!p]
\centering
\includegraphics[angle=0, width=1.0\textwidth]{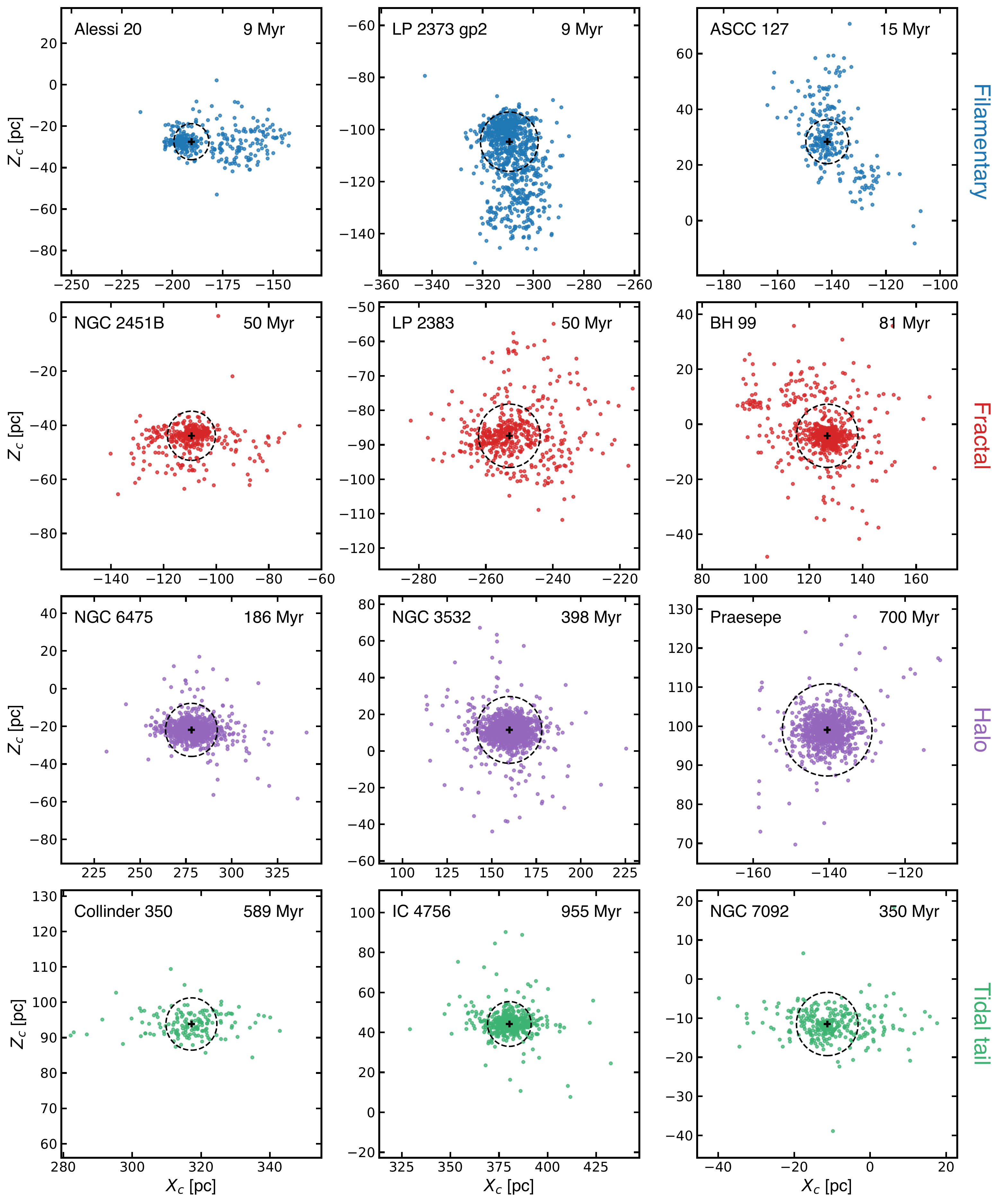}
\caption{
    Example clusters demonstrating the four morphology types in 3D space with filamentary (blue dots), fractal (red), halo (purple) and tidal-tail (green). The name and age of the cluster (group) are indicated in the upper-left and upper-right corners of each panel. The black dashed circle is the tidal radius for each cluster or group. Cluster members shown here all have had their distances corrected.
    }
\label{fig:xz_example}
\end{figure*}

\begin{figure*}[!p]
\centering
\includegraphics[angle=0, width=0.7\textwidth]{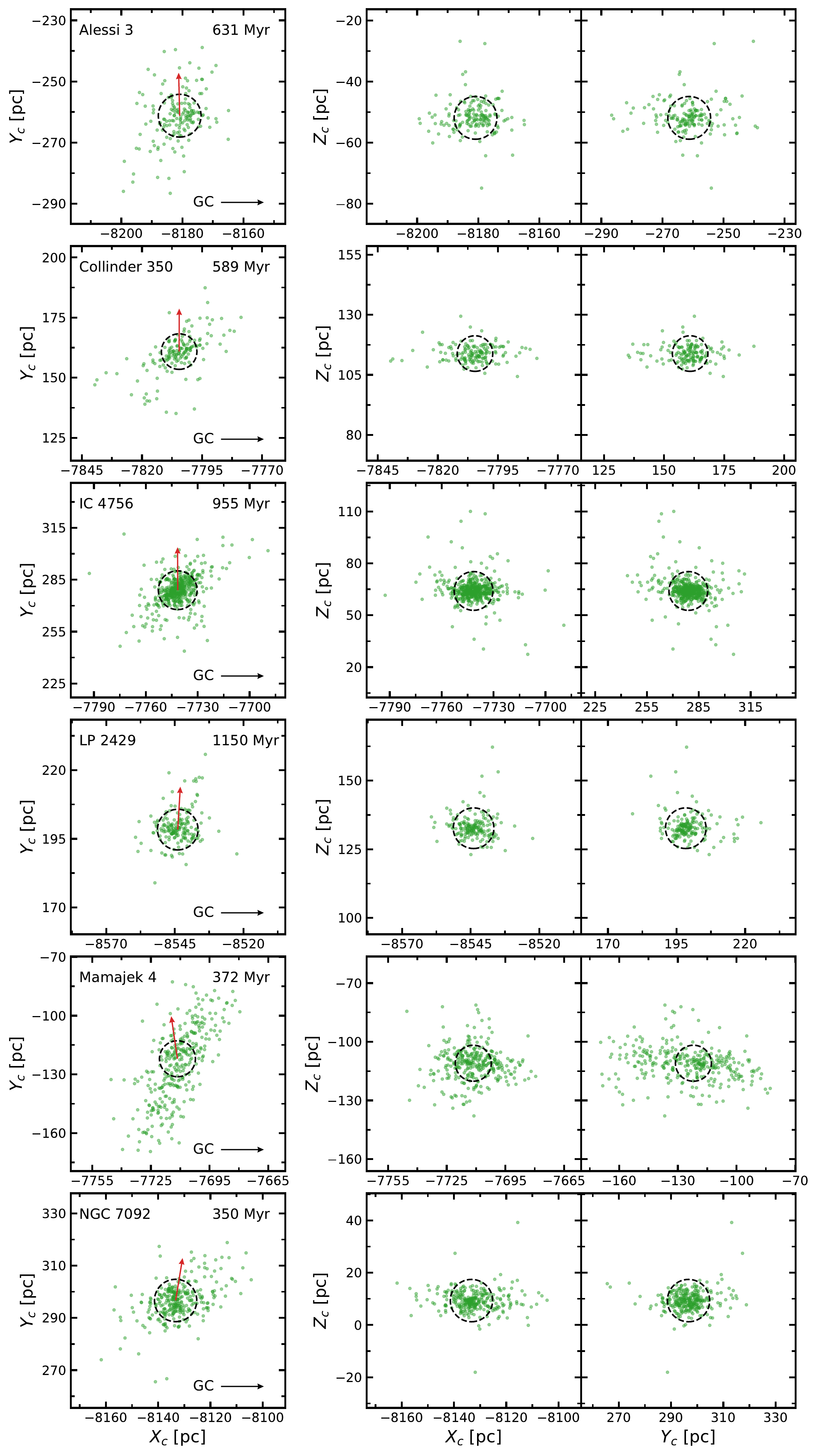}
\caption{
    3D morphology of six tidal-tail clusters. The red arrow indicates the orbital direction of the cluster. The black dashed circle represents the tidal radius of each cluster or group. The names and ages of the clusters are indicated in the upper-left and upper-right corners of each panel. Cluster members shown here all have had their distances corrected.
    }
\label{fig:xyz_tidal}
\end{figure*}

Uni-directional elongated substructures are detected in ten clusters older than 180\,Myr. We consider these substructures as tidal tails and designate them as t-type. Besides four clusters that previously already been reported with tidal tails or early tidal tail structure: Blanco\,1, Coma Berenices, NGC\,2516, and Pleiade \citep{tang2019,Zhang2019,pang2021a,Lodieu19,li2021,Bouma2021}, six more tidal-tail clusters are detected in this study: Alessi\,3, Collinder\,350, IC\,4756, LP\,2429, Mamajek\,4, and NGC\,7092. The 3D spatial distribution of these six clusters is shown in Figure~\ref{fig:xyz_tidal}. The length of the tidal tails in these six clusters ranges from $\sim$20 to $\sim$50\,pc (bottom panels in Figures~\ref{fig:xz_example} and~\ref{fig:xyz_tidal}).  The elongation of the tidal tails is mostly parallel to the Milky Way disk, consistent with \citetalias{pang2021a}, reflecting that the tidal effects in these clusters have successfully overcome the effects of the internal dynamics.

Stars in the tidal tails escape from the cluster through two Lagrangian points \citep[e.g.,][]{kupper2011}. Because of the differential rotation of stars on the Galactic plane, theoretically, stars in the leading tail will orbit faster than the bulk motion of the cluster, while those in the trailing tail will stay behind the cluster as they orbit slower. Such stellar motions in the tails can lead to the observed inclination between tidal tails and the orbital direction of the star cluster (red arrows in the $X$-$Y$ plane in Figure~\ref{fig:xyz_tidal}).

\subsection{Radial Density Profile}\label{sec:radial_density}

To further quantify the four types of clusters hosting morphological substructures, we obtain the radial density profile (RDF) for all 85 target clusters and fit the Elson, Fall, and Freeman (EFF) model \citep{elson1987} to the RDF. The EFF model is characterized by an RDF given by:
\begin{equation}
    \rho(r)=\rho_0\left[1+\left(\frac{r}{a}\right)^2\right]^{-\gamma/2} \quad , 
\end{equation}
where $\rho$ is the number density, and $r$ is the distance of a cluster member to the cluster center. The remaining parameters are fitted; these are the central number density, $\rho_0$; the power-law slope at large radii, $\gamma$; and the scale radius, $a$. A large value of $\gamma$ corresponds to a steep radial distribution, and vice versa. The scale radius $a$ is a measurement of the cluster's physical size. We also compute the core radius, $r_c$, with $a$ and $\gamma$ obtained from fitting the EFF profile using Equation~22 in \citet{elson1987}:
\begin{equation}\label{eq:r_c}
    r_c\approx a \sqrt{2^{2/\gamma}-1} \quad .
\end{equation}

\begin{figure*}[!tbh]
\centering
\includegraphics[angle=0, width=0.95\textwidth]{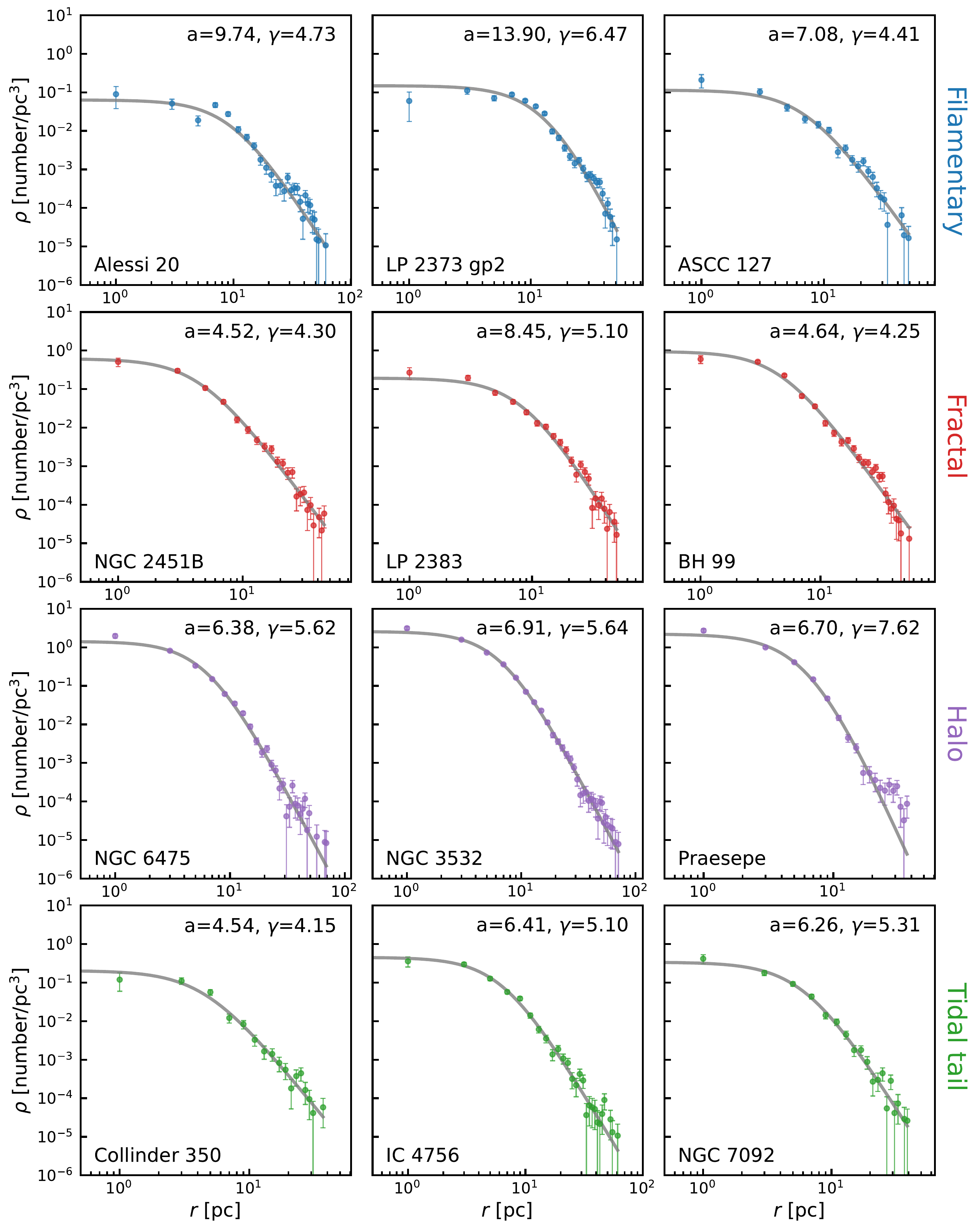}
\caption{
    Radial density profile for example clusters of four morphology types shown in Figure~\ref{fig:xz_example}. The color and symbols are identical to Figure~\ref{fig:xz_example}. $\rho$ is the number density in a cubic parsec and $r$ is the distance of each member to the cluster center. The bin size in each panel is 2\,pc. The error bars are computed with $\rho/\sqrt{N}$, where $N$ is the number of stars in each bin. The grey curve is the fitting result of the Elson, Fall, and Freeman \citep[EFF,][]{elson1987} model to the radial density profile. The best fitted parameters, $a$ and $\gamma$, from the EFF model are indicated in the upper-right corner of each panel.
    }
\label{fig:eff}
\end{figure*}

The fitted results of the EFF model for the example clusters shown in Figure~\ref{fig:xz_example} are displayed in Figure~\ref{fig:eff} (grey curves). Most of the clusters have the highest $\rho$ at the center, except two hierarchical groups, Alessi\,20\,gp1 (f2-type) and UBC\,31\,gp2 (f1-type) having a value for $\rho$ in the innermost 2\,pc that is several times lower than the regions further away from cluster center. These non-centrally concentrated groups are excluded from further analysis, as the EFF profile cannot accurately describe them.

The fitting of EFF profile becomes worse at larger radii ($>$20\,pc), where substructures emerge and have a slight increase in $\rho$ (e.g., filamentary, fractal and halo types in Figure~\ref{fig:eff}). The discrepancy becomes worst in the halo cluster Praesepe (purple), which has an extremely compact profile ($\gamma=7.62$). The $\rho$ value of Praesepe seems to remain constant at radius outside 20\,pc, however, it is a result of low-number statistics (contributed by only 31 stars) so this does not imply any significant feature. This $\rho$ plateau phenomenon becomes less pronounced in tidal-tail clusters, which are at a more advanced disruption state than halo-type. In the tidal-tail clusters, a large number of tidally-stripped stars probably have escaped to large distances from the cluster. Because of this significant mass loss, the density profile consequently becomes the shallowest among the tidal-tail clusters (green). 

\begin{figure*}[!tbh]
\centering
\includegraphics[angle=0, width=1.\textwidth]{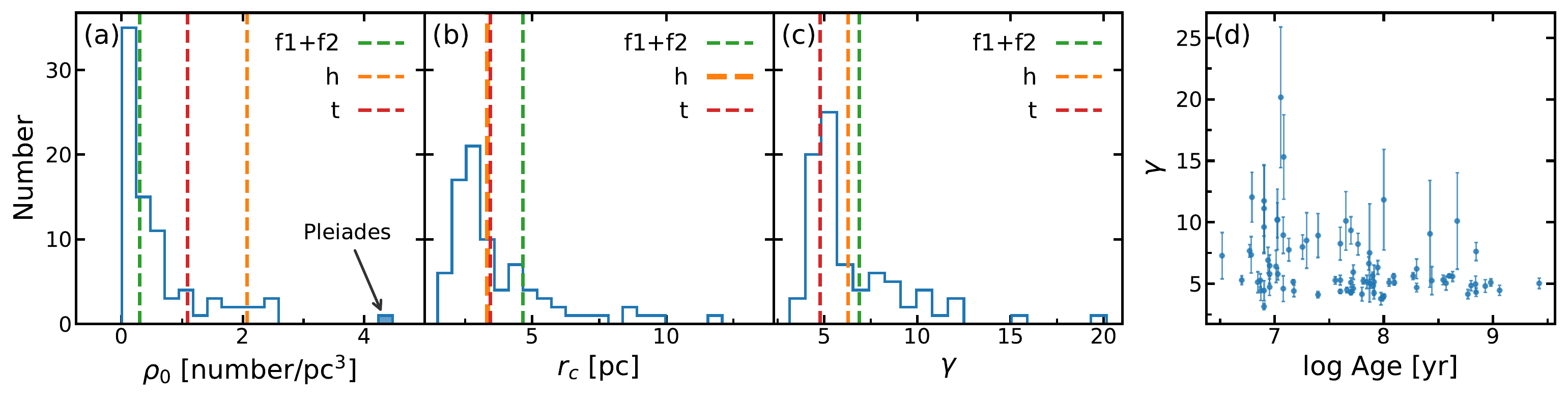}
\caption{
    Histograms of the fitted parameters, $\rho_0$ (central density), $r_c$ (core radius), and $\gamma$ (power-law slope at large radii) for all 85 target clusters (panels a--c), except Group~X, Alessi\,20 gp1, and UBC\,31 gp2. Panel (d) presents the dependence of $\gamma$ on cluster age. $r_c$ is approximated from Equation~\ref{eq:r_c} using the values of $a$ and $\gamma$. The mean value of three types of clusters harboring morphological substructures are presented with dashed vertical lines of different colors. The f1+f2 type is a combination of filamentary (f1) and fractal (f2) clusters (groups). ``h'' and ``t'' represent halo and tidal-tail clusters, respectively. The Pleiades, having the highest $\rho_0$, is highlighted in panel (a).
    }
\label{fig:a_gamma_rc}
\end{figure*}

Considering the similarity in cluster ages of types f1 and f2, we combine these two samples for studying the morphological substructures of young clusters (groups). In Figure~\ref{fig:a_gamma_rc}, we show the distribution of $\rho_0$, $r_c$, and $\gamma$ for three types of clusters (f1+f2, h, and t). Halo clusters with the highest central number density have the highest mean $\rho_0$ (panel (a) in Figure~\ref{fig:a_gamma_rc}). The outlier with the highest  $\rho_0$ value ($>4$\,pc$^{-3}$) is the distinct tidal-tail cluster, the Pleiades.

The average values of $r_c$ is very similar for the h-type and the t-type clusters, while the young f1+f2 type clusters tend to have larger cores. This result is consistent with the earlier study of \cite{tarricq2021}. As clusters grow older, mass segregation manifests. Lower-mass stars tend to obtain higher speeds over time, and migrate to the peripheral area, while massive objects sink to the center \citep[e.g.,][]{hillenbrand1998,pang2013,tang2018}. Over a longer time scale, this process will lead to the shrinking of core radius \citep{heggie2003}.

The value of $\gamma$ decreases from the f1+f2 type to the h-type and to the t-type (Figure~\ref{fig:a_gamma_rc} (c)). A higher value of $\gamma$ indicates more compact profile at large radius. A weak tendency of decreasing $\gamma$ along the increasing cluster age can be seen in panel (d). Young (f1+f2) clusters have a wider range of profiles, both compact and shallow. Older t-type clusters only have small values of $\gamma$ meaning a flat distribution, which should be attributed to their advanced disruption state. Therefore, the value of $\gamma$ can be considered as an indicator of a clusters' dynamical state, which is consistent with the substructure classification based on 3D morphology. Our findings agree with previous results from globular clusters, that tidally-affected clusters have a flatter profile (i.e., a small value of $\gamma$) than the unaffected ones \citep{carballo-bello_outer_2012}.

The unbound stars have been expelled either by stellar encounters or relaxation processes, and form the nearby halo populations. As clusters grow older, more stars will escape, which follow the Galactic shear and form the tidal tails. As a result, clusters with tidal tails have a more shallow and extended density profile than their younger counterparts. Halo clusters and tidal-tail clusters are undergoing similar dynamical disruption processes. Owing to the higher mass and higher density (panel (a) in Figure~\ref{fig:a_gamma_rc}), halo clusters have a longer relaxation time. Therefore, it takes a longer time for the tidal tails to appear.  

\section{Kinematic substructures}\label{sec:kin_feature}

Expansion is often considered as a diagnostic of ongoing cluster dissolution. This signal is seen in both young and old clusters in a number of studies \citep{wright_kinematics_2018,cantat2019b,pang2018,tang2019,pang2021b} using the 3D velocities or the 2D projected motions (the PMs). For old clusters hosting tidal tails, an expansion signal can be interpreted as the presence of the tidal disruption process; however, this is not always true for young clusters (groups). Only young stellar groups showing significant expansion with the speed of members increasing linearly toward the outer region are suggested to be undergoing ``infant'' disruption \citep{wright2018,kuhn2019,pang2021b}. Local expansion in other young groups is a dynamic relics of molecular gas cloud and will not affect the large-scale structure \citep{ward_not_2020}. When dynamical conditions are appropriate, young stellar groups interact and merge \citep{goodwin2004,kuhn_spatial_2015}.  To trace the complexity of stellar motion, we use 3D velocities to search for global dispersion or interacting signatures in our target clusters (groups).

\begin{figure*}[!tbh]
\centering
\includegraphics[angle=0, width=1.01\textwidth]{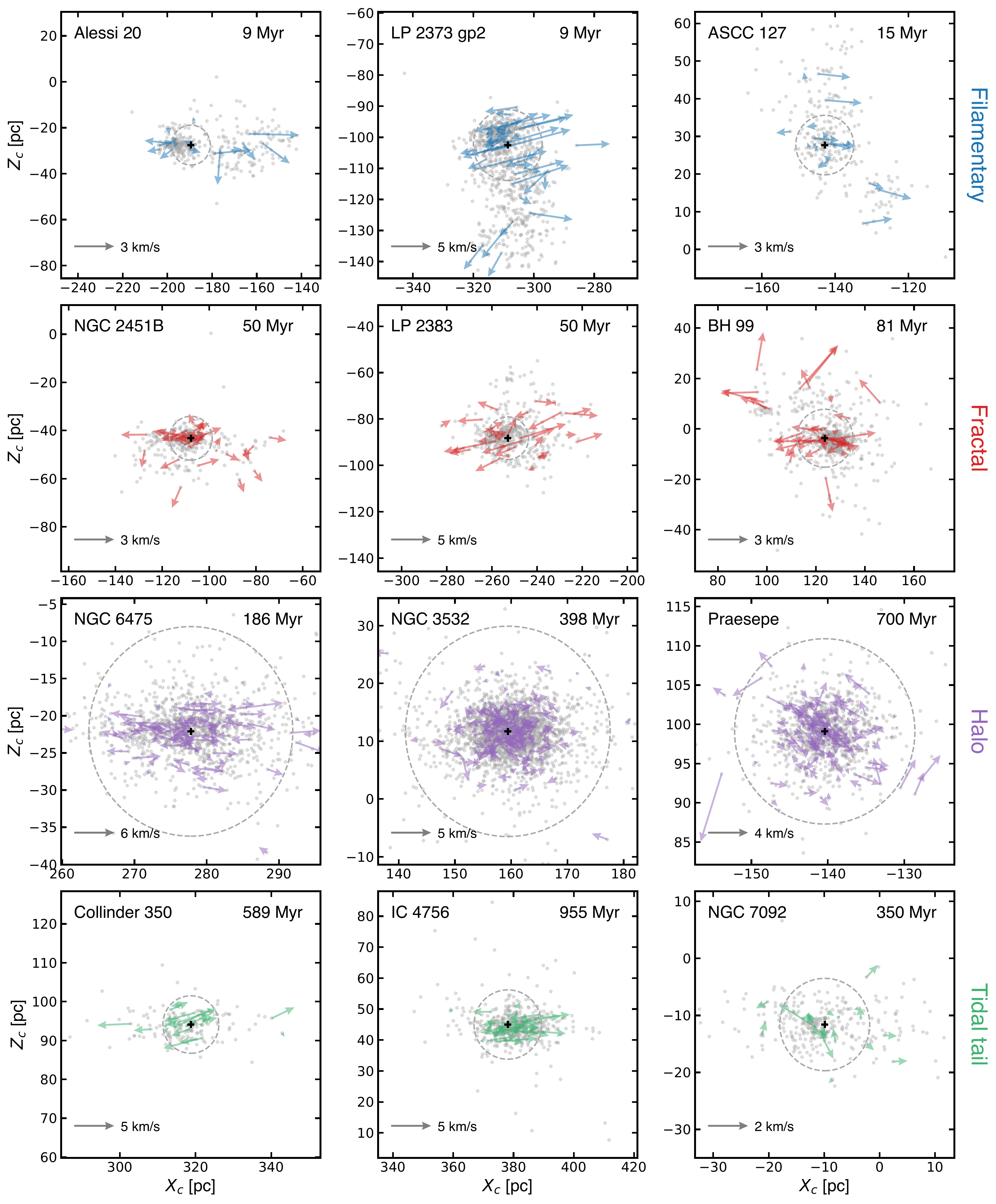}
\caption{
    The relative 3D velocity vectors projected onto the $X$-$Z$ plane, for members of the example clusters in Figure~\ref{fig:xz_example}. The median value of each cluster is taken as the reference. The grey dashed circle represents the tidal radius of each cluster or group. Only 3D velocities within 3 median absolute deviation from the median value are shown. Colors are identical to those in Figure~\ref{fig:xz_example}.
    }
\label{fig:expansion}
\end{figure*}

\subsection{3D Kinematic Substructures}\label{sec:kin_3D}

In Figure~\ref{fig:expansion}, we present the 3D relative velocity (after bulk motion subtracted) computed from PMs and RVs for the example clusters shown in Figure~\ref{fig:xz_example}. The median velocity value of cluster members is taken as the bulk motion of the cluster.  
In the filamentary (f1-type) group (blue arrows), stars of LP\,2373\,gp2 are moving perpendicularly away from the ridgeline of the filamentary structure. A similar trend is also detected in ASCC\,127, but less significant. This kind of ``orthogonal expansion'' seen in young clusters or groups may reveal their destiny of total dispersal. Considering that the uncertainty of Gaia RV might bias the result, we require more accurate kinematic data to confirm the expansion hypothesis. For the fractal (f2-type) clusters (red arrows), simple expansion-like motion is evident in LP\,2383, but stars move in a more complex way in NGC\,2451B and BH\,99. This complexity might be a reflection of turbulence in their parental molecular cloud \citep{elmegreen2000} or an aftermath of a merger event \citep{vazquez-semadeni_hierarchical_2016}. Whether these expansions will impact the young clusters' large-scale structure or not needs to be confirmed with additional high-resolution spectroscopy. With the accuracy of Gaia DR2 RV, we cannot reach a quantitative conclusion. 

The star motions become complex in the bound region (within the tidal radius, grey dashed circle) of halo (h-type) clusters (purple arrows), with both radial expansion and tangential motions, which indicates the simultaneous interplay between two-body relaxation and Galactic tides. Halo clusters are transiting from early tidal disruption toward more advanced disruption stage like the tidal-tail clusters. In the tidal-tail (t-type) clusters (green arrows), the outward motions are inclined to parallel to the Galactic Plane because of the stretch from the Galactic shear.

\subsection{2D Kinematic Substructures}\label{sec:2d_kin}

\begin{figure*}[!tbh]
\centering
\includegraphics[angle=0, width=1.\textwidth]{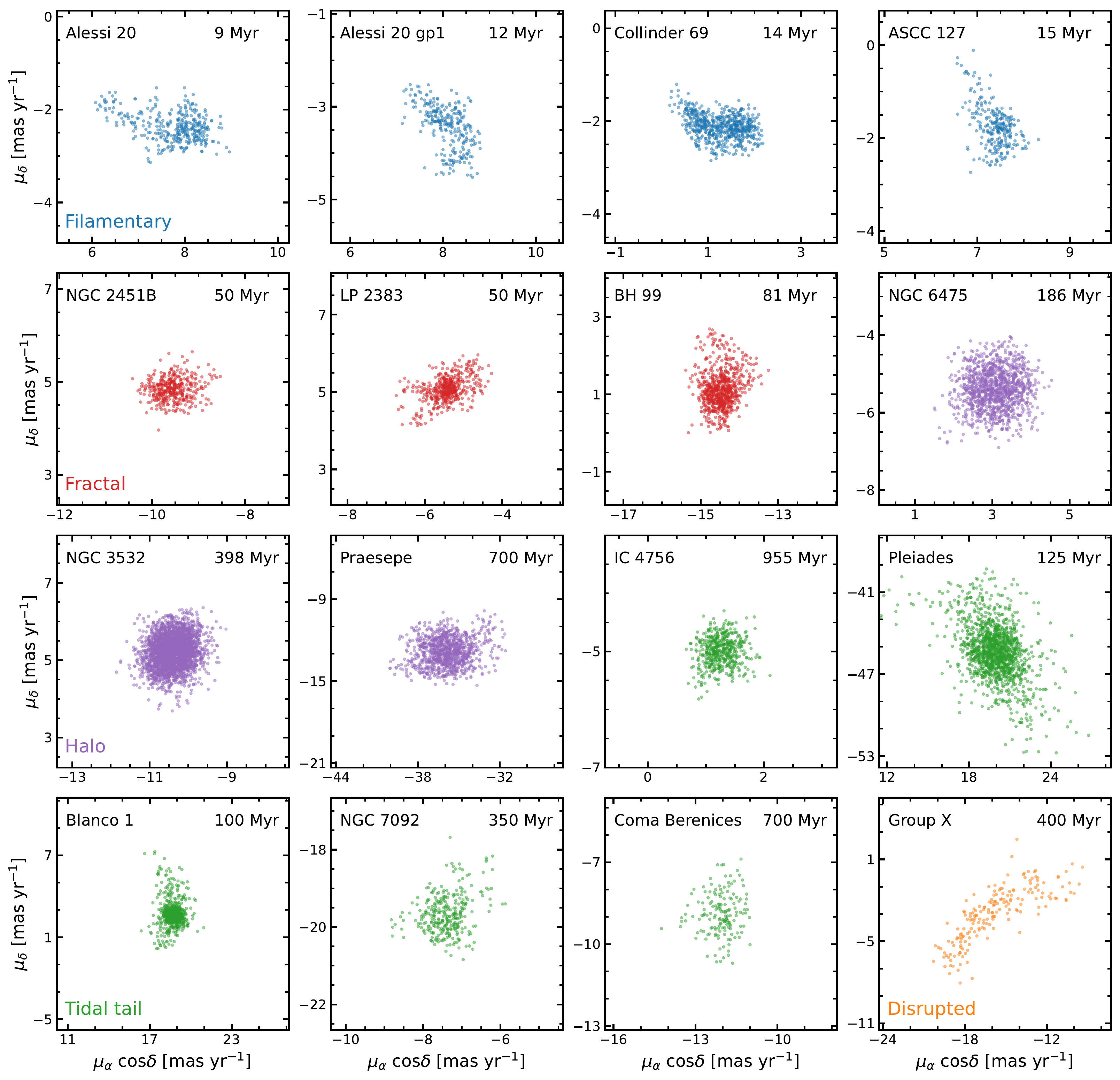}
\caption{
    Proper motion distribution of 16 example clusters of four morphology types: filamentary (blue), fractal (red), halo (purple) and tidal-tail (green), same as Figure~\ref{fig:xz_example}. An extra morphology type, disrupted (orange), for Group\,X is given for a comparison.
    }
\label{fig:PM_mor}
\end{figure*}

\begin{figure*}[!tbh]
\centering
\includegraphics[angle=0, width=1.\textwidth]{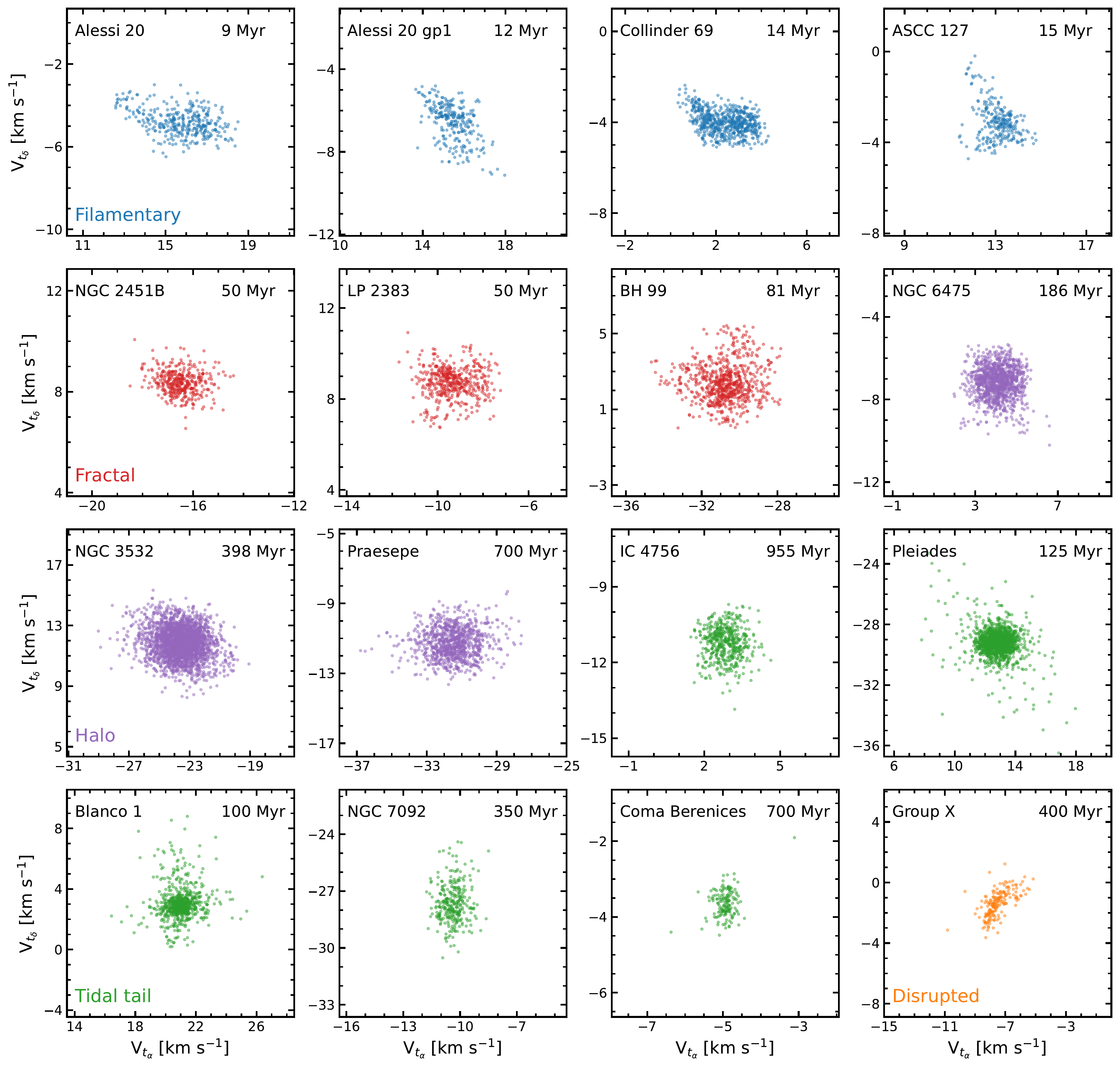}
\caption{ Same as Figure~\ref{fig:PM_mor}, but showing the tangential velocity distribution of 16 example open clusters of four morphology types: filamentary (blue), fractal (red), halo (purple) and tidal-tail (green), same as Figure~\ref{fig:xz_example}. 
    }
\label{fig:tangential_mor}
\end{figure*}

Previous observations have shown that morphological substructures often relate to kinematic substructures in the PM space \citep{wright_cygnus_2016,wright_kinematics_2018,pang2021b}. We present the PM distributions for the example clusters (those in Figure~\ref{fig:xz_example}) in Figure~\ref{fig:PM_mor}. To avoid the projection effect, we convert PMs into tangential velocities ($V_t$;) and show in Figure~\ref{fig:tangential_mor}. Several additional examples of each morphological type are added to explore the difference in the outcome of the kinematic evolution.

The stellar group, Group\,X, is displayed as a reference in both Figures~\ref{fig:PM_mor} and~\ref{fig:tangential_mor} (orange dots) to showcase the PM and $V_t$ distribution for a disrupted stellar group \citep{tang2019}. Like Group\,X, most filamentary (f1-type) clusters have elongated PM and $V_t$ distributions, supporting their unstable and transient dynamical state. This result is consistent with the interpretation of their 3D motions (orthogonal expansion in Section~\ref{sec:kin_3D}). 
As kinematic substructures will be quickly erased by the fast dynamical relaxation or interaction \citep{parker2014}, the elongated kinematic feature we are currently observing in these f1-type young clusters confirm that they are dynamically unevolved. The unbound kinematic substructures are likely relics of the primordial gas kinematics of their unbound parent molecular cloud \citep{heyer2001}. In fractal (f2-type) clusters, the PM and $V_t$ distributions have a higher degree of substructure but less elongation. This difference may be caused by two possibilities: (1) the primordial gas kinematics in f2-type clusters is different from f1-type clusters, owing to different SFE \citep{kruijssen2012}; (2)  dynamical processing has taken place in these clusters so that the inherent kinematics of the molecular gas cloud is affected.

The 2D kinematic distribution of halo (h-type) clusters is more spherical with fine substructures (purple dots in Figure~\ref{fig:PM_mor} and \ref{fig:tangential_mor}). Tail-like substructures appear in the Praesepe, referred to as ``kinematic tails'' \citep{li2021}. The kinematic tails become more significant in tidal-tail (t-type) clusters, e.g., the Pleiades, Blanco\,1, and NGC\,7092. Most stars in the kinematic tails of t-type clusters are located along the tidal tails (Figure~\ref{fig:eff}). Although the Pleiades only shows an early tidal tail substructure outside the tidal radius in spatial space \citep{li2021} that is much shorter than Blanco\,1's tidal tail, 
the remarkably long kinematic tails in both the PM and the $V_t$ distributions may point out the uniqueness of this cluster, e.g., extremely high central density (blue shaded bin in Figure~\ref{fig:a_gamma_rc} (a)). The kinematic substructures in Coma Berenices and IC\,4756, on the other hand, are less intriguing owing to a small number of tidal tail members identified in \citetalias{pang2021a}.

\section{Morphology of hierarchical groups}\label{sec:hiearchy}

\begin{figure*}
\centering
\includegraphics[angle=0, width=1.\textwidth]{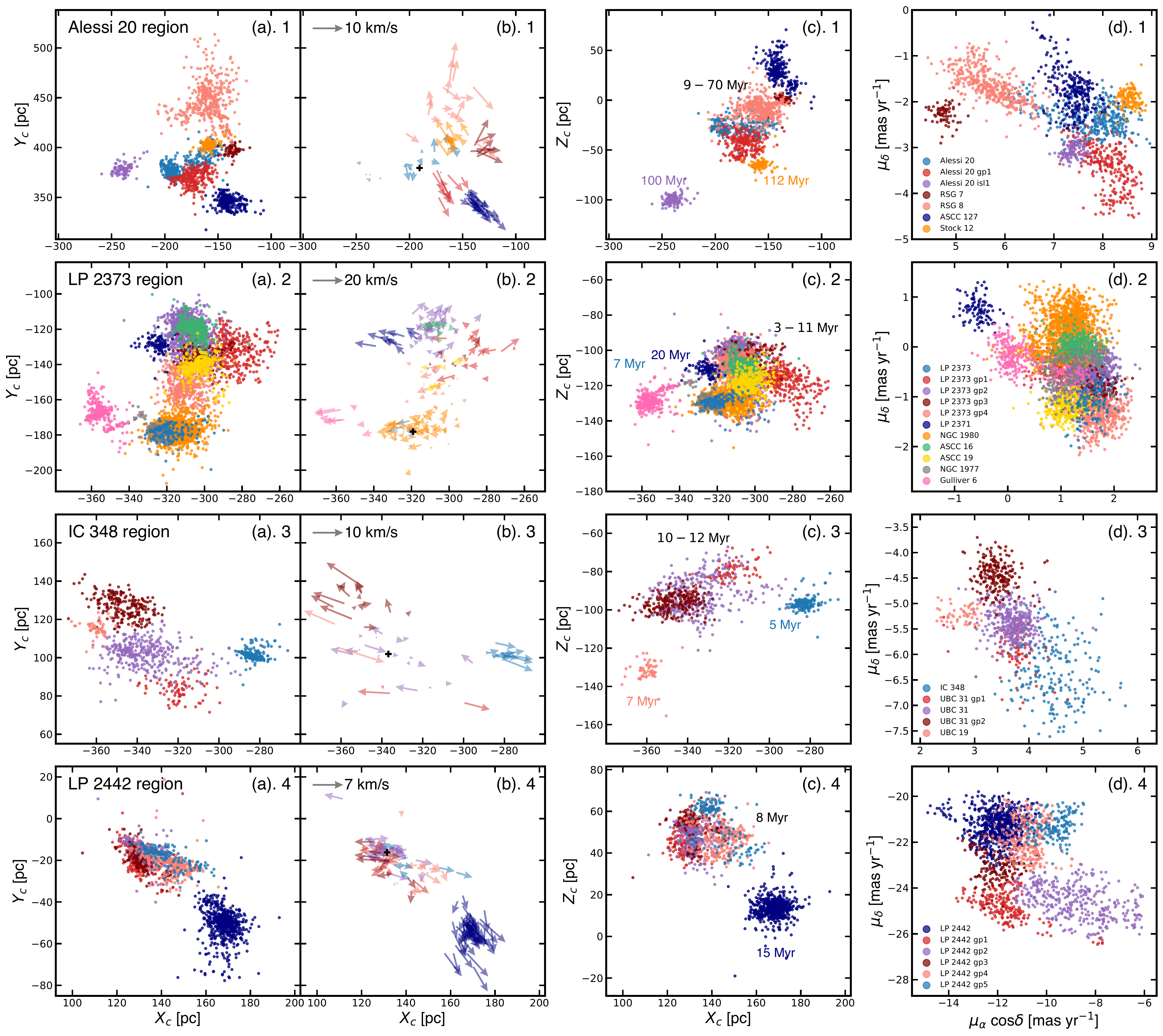}
\caption{
    3D morphology and kinematic feature of four hierarchical clustering regions (panels 1--4). Each color represents one cluster or groups. The (a) and (c) panels display the spatial distribution along $X$-$Y$ and $X$-$Z$ planes in the Heliocentric Cartesian coordinates. (b) panels show velocity vector plots of 3D relative velocity projected onto the $X$-$Z$ plane. The median value of the members in the following clusters is considered as the reference of the corresponding region: Alessi\,20 (panels 1), LP\,2373 (panels 2), UBC\,31 (panels 3), LP\,2442\,gp2 (panels 4). Only 3D velocities within 3 median absolute deviation from the median value are displayed in (b) panels. (d) panels show proper motions distribution of all groups inside each region. An interactive version of panels (a)--(c) is available on: \url{http://3doc-morphology.lowell.edu} under the tab ``Hierarchical Clustering Groups''. Users can choose the desired hierarchical region to display. 
    }
\label{fig:hiearch-group}
\end{figure*}

The most complex morphological and kinematic substructures are seen in the hierarchical clustering groups around four clusters, Alessi\,20, IC\,348, LP\,2373, and LP\,2442. Together with Vela OB2, another sky region found to have hierarchical groups \citep{pang2021b}, we highlight them in the large scale spatial location of solar neighborhood in Figure~\ref{fig:xyz_2D}, and provide a close-in look for the four identified in this study in Figure~\ref{fig:hiearch-group}. Within these four sky regions, we in total identified 29 groups. Ten of the 29 (about one-third) groups are newly discovered, which we name after the target region.

The hierarchical group relating to Alessi\,20, is named Alessi\,20 gp\,1. A new isolated group found around Alessi\,20 is named Alessi\,20 isl\,1. The cluster LP\,2442, first discovered by \citet{liu2019}, also named as UPK\,640 in \citet{Cantat-Gaudin2020} host five hierarchical groups:  LP\,2442 gp\,1--5. These nine groups mentioned above are all newly detected that none are cataloged in \citet{liu2019} or \citet{Cantat-Gaudin2020}.

Six groups are identified in the region of LP\,2373 \citep{liu2019}: LP\,2373 gp\,1--4, ASCC\,16 and ASCC\,19. The latter two are suggested to be a cluster pair in \citet{soubiran_open_2018}. 
LP\,2373 gp\,1 is a newly discovered group that does not shown in \citet{liu2019} and \citet{Cantat-Gaudin2020}. LP\,2372 gp\,2 is the biggest group in the LP\,2372 region with 955 members. With this larger population, LP\,2372 gp\,2 have a fraction of members overlap with that in LP\,2369 \citep[240,][]{liu2019}, LP\,2367 \citep[117,][]{liu2019}, and ASCC\,21 \citep[82,][]{Cantat-Gaudin2020}. As for LP\,2372 gp\,3 (185 members), there are 66 members matches with LP\,2370 \citep[][]{liu2019}. LP\,2372 gp\,4 being the second massive sub-group (511 member), one fifth of its members are matched with UBC\,17a in \citet[183,][]{Cantat-Gaudin2020}. 

The majority of the hierarchical groups are younger than 30\,Myr. They are either filamentary (f1) or fractal (f2) type (Section~\ref{sec:class}). We plot their 3D spatial distribution, 3D velocity, and PM distribution in Figure~\ref{fig:hiearch-group}. The morphology of spatial substructures in these four regions varies significantly (Figure~\ref{fig:hiearch-group} panels (a) and (c)). Consistent with morphological substructures, a higher degree of kinematic substructures (Figure~\ref{fig:hiearch-group} panels (d)) are found in these hierarchical groups. 
We discuss the unique features of each hierarchical clustering region in the following sections. 
 
\subsection{Dispersing Filament Network: Alessi\,20 and LP\,2373 Regions}\label{alessi20}

Four groups, Alessi\,20, Alessi gp\,1, RSG\,8, and ASCC\,127 connect in space, resembling filamentary networks (panels (a).1 and (c).1). At the same time, the PM morphology of these four groups is elongated and overlaps at the location of the youngest group, Alessi\,20 ($\sim$9\,Myr, panel (d).1). This filament network in the Alessi\,20 region looks very similar to those filament-hub or filament-ridge systems in the observed molecular clouds \citep{schneider2010,trevino-morales2019}. This is direct evidence of the inherent morphological and kinematic substructure from molecular clouds. Considering the age spread among groups, star formation probably propagates along the filaments. The oldest group Alessi\,20 isl\,1 ($\sim$100\,Myr) is the older generation in this star formation region.

In the relative velocity vector diagram (panel (b).1), an obvious expansion signature is detected with stars moving away from the reference center, Alessi\,20. Such expansion probably disperse the filament within a short time. In the cluster pair, RSG\,7 ($\sim$70\,Myr) and RSG\,8 ($\sim$18\,Myr), only a fraction of stars in RSG\,8 moving toward RSG\,7. Their chance of interaction will be much lower than the colliding pair Collinder\,350 and IC\,4665 \citep{piatti2022}.

The filament network phenomenon becomes weaker in the LP\,2373 region as the morphology of groups turns into more fractal ((a).2 and (c).2). The hierarchical groups, LP\,2373 gp\,1--4, in this region largely overlap around the youngest group LP\,2373 ($\sim$5\,Myr) in both the spatial space (blue points in (a).2 and (c).2) and in the PM space (blue points (d).2). The overall ``orthogonal expansion'' observed in panel (b).2 might reshape the morphology of groups in this region and eventually disrupt the filamentary substructures.

The dispersing signature detected in both Alessi\,20 and LP\,2373 regions probably is attributed to a low SFE (less than 1/3) at their birthplace inside the parental clouds \citep{kruijssen2012}. Therefore, these groups cannot survive the residual gas expulsion \citep{baumgardt2007}. Despite the young age of Alessi\,20 ($\sim$9--18\,Myr) and LP\,2373 groups ($\sim$4--13\,Myr), we are probably witnessing an ongoing ``filament network dissolution''. Tidal effect may start to stretch these groups, however, we cannot distinguish tidal induced substructure from the primordial filamentary substructure in the current analysis.

\subsection{In-falling Motions: IC\,348 Region}\label{ic348}

All five groups of stars in IC\,348 regions distribute along with filamentary structures ((a).3 and (c).3). Their PM distribution ((d).3) is elongated and adjacent to one another. Unexpectedly, IC\,348, the youngest group, is much dispersed in the PM diagram, opposite to its concentrated spatial morphology. A tangential motion is observed among members in the hierarchical groups ((b).3), with most stars in UBC\,31, UBC\,31 gp\,1 and UBC\,31 gp\,2 moving along the filament. This motion is similar to the infalling signature in the filaments in Cygnus X and Orion Nebula \citep{schneider2010,kounkel_gravitational_2021}. The stellar groups in IC\,348 regions are surrounded by big molecular clouds \citep{gutermuth_spitzer_2009}. The attraction of another more profound potential might trigger this in-falling.

\subsection{Well Mixed Structure: LP\,2442 Region}\label{lp2442}

Unlike the other three regions, the feature in LP\,2442 is more complex. Five coeval hierarchical subgroups are well mixed in the 3D positional space ((a).4 and (c).4). They are, however, easier to distinguish in the PM diagram ((d).4). Their adjacency in the PM distribution indicates the kinematic coherence among five subgroups. These five hierarchical subgroups might have experienced merging events for the well-mixed spatial configuration. The 3D motions of stars in these five subgroups (relative to LP\,2442 gp\,2, (b).4) are both along radially inward and outward directions, different from the expansion signature seen in Alessi\,20 and LP\,2373 region. Only when these subgroups are sufficiently dynamically cool will they interact to form a dense star cluster by hierarchical mergers \citep{allison2009,fujii2021}. Once the virial ratio switches to supervirial, these subgroups will depart from each other. Because of this complexity, it is unclear whether these merging motions seen in the LP\,2442 region will lead to a dense cluster.

The complex kinematic substructures in these four hierarchical regions indicate that they are young in stellar and dynamical evolution. Dynamical relaxation haves no time to shape the spatial morphology of these groups and has no chance to erase the primordial kinematics. Therefore, the observed filamentary substructures in these four regions are \emph{in situ}, as birthmarks from molecular clouds \citep{ballone2020}. Our results agree with previous finding on the ``hierarchical star formation'' scenario \citep{kruijssen2012,wright2020}. At the same time, we show a large variety of morphological and kinematic substructures in young stellar regions. Future high precision of kinematic data are desperately needed to quantify the dynamical state of these hierarchical groups.

\section{Morphology Inside Tidal Radius}\label{sec:mor_tidal}

As demonstrated in Section~\ref{sec:member_reliablity}, cluster members identified within the tidal radius are most stringent and least affected by the contamination rate. In this section, we analyze the shape of all our target clusters within the bound region following the method developed in \citetalias{pang2021a} by performing the ellipsoid fitting\footnote{\url{https://github.com/marksemple/pyEllipsoid_Fit}}. 
In \citetalias{pang2021a}, we found that, in general, the spatial distribution of member stars within the tidal radius of open clusters can be well described by oblate spheroids, prolate spheroids, or triaxial ellipsoids. 
During the ellipsoid fitting, we first center the fitted ellipsoid at the median position of all bound members, where we considered the cluster center. For the fitted semi-axes of the ellipsoid, $a$, $b$, and $c$, we make $a$ as the semi-major axis, $b$ the semi-intermediate axis, and $c$ the semi-minor axis. We use the lengths of the semi-axes $a$, $b$, $c$, and axis ratios $b/a$ and $c/a$ to describe the morphology of the clusters. The direction of the semi-major axis $a$ is defined as the direction of elongation of the fitted cluster. Smaller values of both the axis ratios, $b/a$ and $c/a$, indicate a more elongated structure. The fitted values of the morphological parameters ($a$, $b$, $c$, $b/a$ and $c/a$) are listed in Table~\ref{tab:morph_kin}. We also compute for each cluster the angle $\theta$ between the direction of $a$ and the Galactic plane, and the angle $\phi$ between the direction of $a$ and the line-of-sight. The values of these two angles are also given in Table~\ref{tab:morph_kin}. The morphological parameters of the 12 clusters from \citetalias{pang2021a} are taken from Table~3 in \citetalias{pang2021a}.

\begin{figure*}
\centering
\includegraphics[angle=0, width=0.85\textwidth]{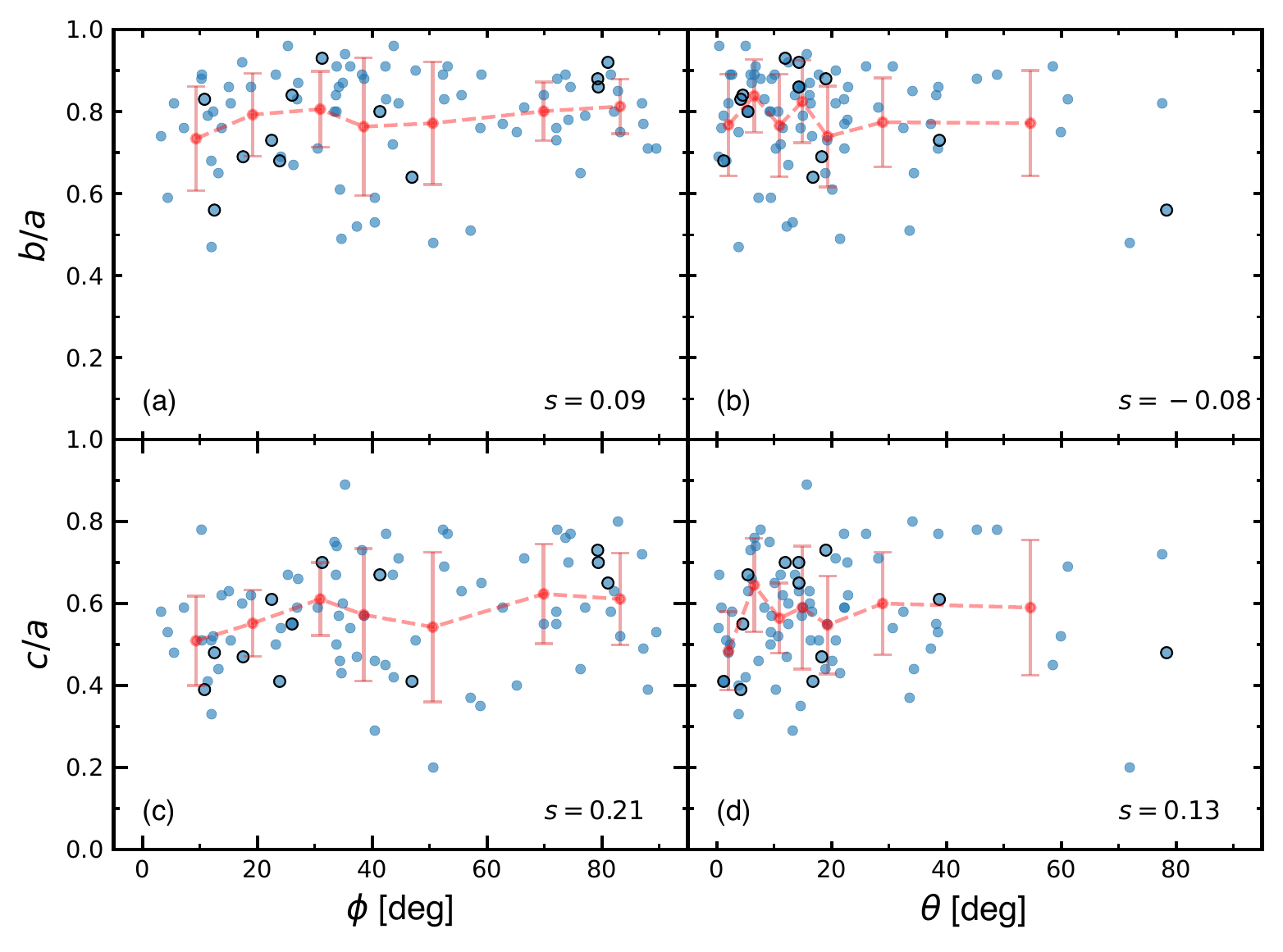}
\caption{
    Dependence of axis ratio of the fitted ellipsoid on the angle $\phi$ and $\theta$. $\phi$ is the angle between the semi-major axis $a$ and the direction of the line-of-sight. $\theta$ is the angle between the semi-major axis $a$ and the Galactic Plane. Each blue dot represent an single cluster. The red dots and error bars are the mean and standard deviation in each bin. To ensure a statistical significance, we keep a constant number of 12 clusters in each bin. The 12 clusters with parameters from \citetalias{pang2021a} are highlighted with black open circles. The Spearman’s rank correlation coefficients ($s$) are shown in each panel. An $s$ value close to 1 indicates strong correlation between two variables, and close to 0 indicates no correlation.
    }
\label{fig:axis_ratio_angle}
\end{figure*}

As can be seen in Figure~\ref{fig:axis_ratio_angle}, there is no dependence of the ratio $b/a$ on $\phi$ (the angle between the elongated direction and the line-of-sight direction) with the Spearman coefficient: $s=0.09$, and $c/a$ on $\phi$ with $s=0.21$. The slightly larger value of $s=0.21$ in $c/a$ vs. $\phi$ does not truly imply any correlation. This evidence reassures that the distance correction is appropriate for members within the tidal radius. Surprisingly, the $c/a$ ratio is also independent of $\theta$ ($s=0.13$), the angle between the direction of elongation and the Galactic Plane. As the long-term effect of the Galactic tides on clusters is to make their members show elongation parallel to the Galactic plane, this ``lie down'' effect is shown as the smaller $\theta$ goes, the smaller the $c/a$ ratio. This trend is not seen in our cluster samples, we find that filamentary structures (small $c/a$) in young clusters tend to have a random angle relative to the Galactic plane (f1-type in Figure~\ref{fig:xz_example}, Figure~\ref{fig:relation-age} (b)), which ``successfully''  dilutes the anticipated effect produced by tidal field.

Based on the axis ratio, we define three kinds of shape describing the spatial distributions of stars inside tidal radius, based on the following criteria:

\begin{enumerate}
    \item Type~I: oblate spheroid:
        \begin{equation}
           \frac{c}{a}<\frac{b}{a},\quad \frac{b}{a}\ge0.8,\quad \frac{c}{a}<0.6
        \end{equation}
    \item Type~II: prolate spheroid:
        \begin{equation}
        \left|\frac{c}{a}-\frac{b}{a}\right|<0.1,\quad \frac{b}{a}<0.6, \quad \frac{c}{a}<0.6
        \end{equation}
    \item Type~III: triaxial ellipsoid:
        \begin{equation}
         \left|\frac{c}{a}-\frac{b}{a}\right|\ge0.2
        \end{equation}
        \\
\end{enumerate}

The shapes of stellar populations inside the tidal radius give 14 clusters that resemble an oblate ellipsoid, three as a prolate ellipsoid, and seven as a triaxial ellipsoid. Pleiades, the cluster with a prominent kinematics tail (Figures~\ref{fig:PM_mor} and~\ref{fig:tangential_mor}), exhibits a oblate shape. The most massive cluster in our sample, NGC\,3532, resembles a triaxial ellipsoid. The tidal-tail cluster, Collinder\,350 is oblate, with its $a$ axis perfectly align with the Galactic Plane. In all the hierarchical clustering groups, the LP\,2442 region hosts the largest variety of morphologies. Three groups are oblate (LP\,2442, LP\,2442 gp\,1, LP\,2442 gp\,4), two triaxial (LP\,2442 gp\,2, LP\,2442 gp\,5), and one prolate (LP\,2442 gp\,3).

The diversity of cluster morphology inside the tidal radius reveals different dynamical state among clusters. The discussion of individual cluster dynamical state requires a higher precision kinematic data and therefore is beyond the scope of this paper. However, the parameterization of 3D positional distribution of clusters do allow us to investigate the dependence of cluster morphology (e.g., axis ratio $c/a$) on other structural parameters (e.g., $r_{\rm h}$) and properties (e.g., age), paving the way to understand how the internal and external dynamical processes reshape the cluster's spatial distribution overtime.

\startlongtable
\begin{deluxetable*}{L R RRR RR RR R}
\tablecaption{Morphological Parameters of 72 Target Clusters. \label{tab:morph_kin}
}
\tabletypesize{\footnotesize}
\tablehead{
	 \colhead{Cluster}  & \colhead{$\sigma_{dis}$}  &
	 \colhead{$a$}      & \colhead{$b$}             & \colhead{$c$} & 
	 \colhead{$b/a$}   & \colhead{$c/a$}    &          
	 \colhead{$\theta$}  & \colhead{$\phi$} & \colhead{Type}    \\
	 \colhead{}                 &
	 \multicolumn{4}{c}{(pc)}   &
	 \colhead{}                 & \colhead{}    &
	 \multicolumn{2}{c}{(degrees)}              & \colhead{}    \\
	 \cline{2-5} \cline{8-9} 
	 \colhead{(1)} & \colhead{(2)} & \colhead{(3)} &
	 \colhead{(4)} & \colhead{(5)} &
	 \colhead{(6)} & \colhead{(7)} &
	 \colhead{(8)} & \colhead{(9)} &
	 \colhead{(10)}
	 }
\startdata
\rm Alessi\,3	&	1.9	&	5.76	\pm	1.33	&	4.53	\pm	0.21	&	3.4	\pm	0.26	&	0.79	\pm	0.19	&	0.59	\pm	0.14	&	15.02	&	77.1	&	\rm III	\\
\rm Alessi\,5	&	2.7	&	5.29	\pm	2.87	&	4.85	\pm	0.38	&	3.17	\pm	0.46	&	0.92	\pm	0.5	&	0.6	\pm	0.34	&	12.49	&	17.37	&	\rm III	\\
\rm Alessi\,9	&	1.7	&	6.34	\pm	1.02	&	3.1	\pm	0.16	&	2.73	\pm	0.21	&	0.49	\pm	0.08	&	0.43	\pm	0.08	&	21.49	&	34.66	&	\rm II	\\
\rm Alessi\,20	&	3.0	&	6.35	\pm	3.47	&	4.52	\pm	0.44	&	3.73	\pm	0.51	&	0.71	\pm	0.39	&	0.59	\pm	0.33	&	22.25	&	30.53	&		\\
\rm Alessi\,20\,gp1$^a$	&	3.0	&	6.29	\pm	3.42	&	5.38	\pm	0.43	&	3.97	\pm	0.52	&	0.86	\pm	0.47	&	0.63	\pm	0.35	&	14.32	&	15.04	&	\rm III	\\
\rm Alessi\,20\,isl1$^a$	&	3.4	&	6.0	\pm	4.4	&	4.25	\pm	0.51	&	3.16	\pm	0.59	&	0.71	\pm	0.53	&	0.53	\pm	0.4	&	38.54	&	89.49	&		\\
\rm Alessi\,24	&	4.0	&	5.41	\pm	6.41	&	4.42	\pm	0.65	&	2.75	\pm	0.73	&	0.82	\pm	0.97	&	0.51	\pm	0.62	&	16.37	&	15.38	&	\rm I	\\
\rm Alessi\,62	&	3.8	&	6.86	\pm	5.48	&	4.08	\pm	0.59	&	3.16	\pm	0.65	&	0.59	\pm	0.48	&	0.46	\pm	0.38	&	7.29	&	40.47	&		\\
\rm ASCC\,16	&	2.4	&	6.41	\pm	2.13	&	4.93	\pm	0.31	&	3.81	\pm	0.36	&	0.77	\pm	0.26	&	0.59	\pm	0.21	&	22.25	&	62.74	&		\\
\rm ASCC\,19	&	2.6	&	6.95	\pm	2.65	&	5.11	\pm	0.36	&	4.03	\pm	0.42	&	0.74	\pm	0.28	&	0.58	\pm	0.23	&	16.61	&	3.24	&		\\
\rm ASCC\,58	&	4.0	&	6.42	\pm	6.11	&	5.42	\pm	0.63	&	4.32	\pm	0.69	&	0.84	\pm	0.81	&	0.67	\pm	0.65	&	13.64	&	33.7	&		\\
\rm ASCC\,105	&	3.2	&	4.56	\pm	4.06	&	3.97	\pm	0.5	&	2.75	\pm	0.57	&	0.87	\pm	0.78	&	0.6	\pm	0.55	&	16.22	&	34.89	&	\rm III	\\
\rm ASCC\,127	&	2.5	&	5.96	\pm	2.37	&	5.15	\pm	0.33	&	3.7	\pm	0.4	&	0.86	\pm	0.35	&	0.62	\pm	0.26	&	22.88	&	18.9	&	\rm III	\\
\rm BH\,99	&	3.8	&	7.69	\pm	5.64	&	5.83	\pm	0.61	&	4.77	\pm	0.67	&	0.76	\pm	0.56	&	0.62	\pm	0.46	&	11.51	&	13.79	&		\\
\rm BH\,164	&	3.6	&	6.97	\pm	5.01	&	5.26	\pm	0.56	&	2.78	\pm	0.63	&	0.75	\pm	0.55	&	0.4	\pm	0.3	&	3.82	&	65.19	&	\rm III	\\
\rm Collinder\,69	&	3.1	&	7.64	\pm	3.87	&	5.94	\pm	0.48	&	5.34	\pm	0.55	&	0.78	\pm	0.4	&	0.7	\pm	0.36	&	22.77	&	74.18	&		\\
\rm Collinder\,135$^b$	&	2.2	&	7.17	\pm	1.77	&	6.42	\pm	0.27	&	3.67	\pm	0.31	&	0.9	\pm	0.22	&	0.51	\pm	0.13	&	20.74	&	47.57	&	\rm I	\\
\rm Collinder\,140	&	2.8	&	6.64	\pm	3.09	&	5.9	\pm	0.42	&	3.29	\pm	0.46	&	0.89	\pm	0.42	&	0.5	\pm	0.24	&	2.36	&	23.23	&	\rm I	\\
\rm Collinder\,350	&	2.6	&	5.3	\pm	2.68	&	4.68	\pm	0.37	&	4.11	\pm	0.43	&	0.88	\pm	0.45	&	0.78	\pm	0.4	&	7.65	&	10.28	&		\\
\rm Gulliver\,6	&	3.1	&	5.54	\pm	3.67	&	4.58	\pm	0.48	&	3.29	\pm	0.54	&	0.83	\pm	0.55	&	0.59	\pm	0.4	&	8.31	&	26.93	&		\\
\rm Gulliver\,21	&	3.7	&	5.66	\pm	5.29	&	3.47	\pm	0.57	&	2.59	\pm	0.65	&	0.61	\pm	0.58	&	0.46	\pm	0.44	&	20.12	&	34.4	&		\\
\rm Huluwa\,1$^b$	&	2.5	&	9.56	\pm	2.47	&	9.17	\pm	0.36	&	6.42	\pm	0.41	&	0.96	\pm	0.25	&	0.67	\pm	0.18	&	0.44	&	25.33	&	\rm III	\\
\rm Huluwa\,2$^b$	&	2.9	&	9.21	\pm	3.09	&	7.68	\pm	0.4	&	6.38	\pm	0.46	&	0.83	\pm	0.28	&	0.69	\pm	0.24	&	61.15	&	52.55	&		\\
\rm Huluwa\,3$^b$	&	2.6	&	7.3	\pm	2.57	&	6.51	\pm	0.36	&	5.53	\pm	0.41	&	0.89	\pm	0.32	&	0.76	\pm	0.27	&	6.58	&	73.68	&		\\
\rm Huluwa\,4$^b$	&	2.5	&	6.33	\pm	2.29	&	5.17	\pm	0.32	&	4.53	\pm	0.39	&	0.82	\pm	0.3	&	0.72	\pm	0.27	&	77.58	&	87.0	&		\\
\rm Huluwa\,5$^b$	&	2.1	&	4.55	\pm	1.72	&	3.62	\pm	0.25	&	2.35	\pm	0.31	&	0.8	\pm	0.31	&	0.52	\pm	0.21	&	10.7	&	12.33	&	\rm I	\\
\rm IC\,348	&	2.6	&	4.82	\pm	2.59	&	3.15	\pm	0.36	&	2.13	\pm	0.42	&	0.65	\pm	0.36	&	0.44	\pm	0.25	&	18.94	&	13.23	&	\rm III	\\
\rm IC\,4756	&	3.0	&	8.44	\pm	3.58	&	6.4	\pm	0.45	&	5.0	\pm	0.51	&	0.76	\pm	0.33	&	0.59	\pm	0.26	&	0.83	&	7.22	&		\\
\rm LP\,2371	&	2.8	&	6.25	\pm	3.0	&	4.08	\pm	0.42	&	2.75	\pm	0.46	&	0.65	\pm	0.32	&	0.44	\pm	0.22	&	34.35	&	76.29	&		\\
\rm LP\,2373	&	2.9	&	5.64	\pm	3.21	&	5.39	\pm	0.41	&	2.37	\pm	0.5	&	0.96	\pm	0.55	&	0.42	\pm	0.25	&	5.04	&	43.75	&	\rm I	\\
\rm LP\,2373\,gp1$^a$	&	2.5	&	6.63	\pm	2.42	&	5.52	\pm	0.34	&	5.08	\pm	0.4	&	0.83	\pm	0.31	&	0.77	\pm	0.29	&	22.18	&	42.43	&		\\
\rm LP\,2373\,gp2	&	2.5	&	8.53	\pm	2.32	&	7.3	\pm	0.33	&	6.54	\pm	0.39	&	0.86	\pm	0.24	&	0.77	\pm	0.21	&	38.6	&	74.55	&		\\
\rm LP\,2373\,gp3	&	2.5	&	5.12	\pm	2.49	&	4.51	\pm	0.36	&	4.0	\pm	0.41	&	0.88	\pm	0.43	&	0.78	\pm	0.39	&	45.31	&	72.23	&		\\
\rm LP\,2373\,gp4	&	2.5	&	6.78	\pm	2.42	&	6.37	\pm	0.33	&	6.03	\pm	0.41	&	0.94	\pm	0.34	&	0.89	\pm	0.32	&	15.69	&	35.28	&		\\
\rm LP\,2383	&	3.7	&	8.18	\pm	5.4	&	4.85	\pm	0.59	&	4.32	\pm	0.67	&	0.59	\pm	0.4	&	0.53	\pm	0.36	&	9.44	&	4.4	&	\rm II	\\
\rm LP\,2388	&	4.4	&	5.65	\pm	7.44	&	5.15	\pm	0.69	&	3.04	\pm	0.76	&	0.91	\pm	1.2	&	0.54	\pm	0.72	&	30.66	&	36.2	&	\rm I	\\
\rm LP\,2428	&	3.4	&	5.18	\pm	4.54	&	4.34	\pm	0.52	&	3.28	\pm	0.6	&	0.84	\pm	0.74	&	0.63	\pm	0.57	&	16.17	&	55.57	&	\rm III	\\
\rm LP\,2429	&	3.5	&	8.67	\pm	4.84	&	4.63	\pm	0.54	&	2.51	\pm	0.64	&	0.53	\pm	0.3	&	0.29	\pm	0.18	&	13.24	&	40.47	&	\rm III	\\
\rm LP\,2439	&	2.2	&	5.16	\pm	1.86	&	4.57	\pm	0.28	&	4.05	\pm	0.33	&	0.89	\pm	0.32	&	0.78	\pm	0.29	&	48.83	&	52.34	&		\\
\rm LP\,2441	&	2.1	&	6.57	\pm	1.63	&	5.35	\pm	0.25	&	4.65	\pm	0.31	&	0.81	\pm	0.21	&	0.71	\pm	0.18	&	28.15	&	66.49	&		\\
\rm LP\,2442	&	1.4	&	6.93	\pm	0.64	&	6.07	\pm	0.11	&	3.96	\pm	0.15	&	0.88	\pm	0.08	&	0.57	\pm	0.06	&	9.57	&	38.62	&	\rm I	\\
\rm LP\,2442\,gp1$^a$	&	0.7	&	5.34	\pm	0.12	&	4.48	\pm	0.06	&	2.92	\pm	0.05	&	0.84	\pm	0.02	&	0.55	\pm	0.02	&	38.23	&	69.89	&	\rm I	\\
\rm LP\,2442\,gp2$^a$	&	0.8	&	6.87	\pm	0.19	&	5.15	\pm	0.06	&	3.56	\pm	0.06	&	0.75	\pm	0.02	&	0.52	\pm	0.02	&	59.94	&	83.21	&	\rm III	\\
\rm LP\,2442\,gp3$^a$	&	0.8	&	5.55	\pm	0.16	&	2.86	\pm	0.06	&	2.6	\pm	0.06	&	0.52	\pm	0.02	&	0.47	\pm	0.02	&	12.2	&	37.34	&	\rm II	\\
\rm LP\,2442\,gp4$^a$	&	1.2	&	5.38	\pm	0.44	&	4.91	\pm	0.08	&	2.42	\pm	0.1	&	0.91	\pm	0.08	&	0.45	\pm	0.04	&	58.54	&	42.31	&	\rm I	\\
\rm LP\,2442\,gp5$^a$	&	0.8	&	8.64	\pm	0.17	&	4.12	\pm	0.07	&	1.69	\pm	0.06	&	0.48	\pm	0.01	&	0.2	\pm	0.01	&	71.93	&	50.65	&	\rm III	\\
\rm Mamajek\,4	&	3.0	&	7.07	\pm	3.45	&	5.98	\pm	0.46	&	5.63	\pm	0.52	&	0.85	\pm	0.42	&	0.8	\pm	0.4	&	34.1	&	82.81	&		\\
\rm NGC\,1901	&	2.3	&	5.0	\pm	1.96	&	4.44	\pm	0.29	&	3.25	\pm	0.35	&	0.89	\pm	0.35	&	0.65	\pm	0.26	&	10.1	&	59.03	&	\rm III	\\
\rm NGC\,1977	&	2.4	&	4.97	\pm	2.24	&	3.83	\pm	0.3	&	2.46	\pm	0.38	&	0.77	\pm	0.35	&	0.49	\pm	0.24	&	37.3	&	87.23	&		\\
\rm NGC\,1980	&	2.7	&	10.52	\pm	2.66	&	7.97	\pm	0.37	&	3.67	\pm	0.42	&	0.76	\pm	0.19	&	0.35	\pm	0.1	&	14.61	&	58.85	&	\rm III	\\
\rm NGC\,2451B	&	2.7	&	6.0	\pm	2.89	&	5.36	\pm	0.38	&	4.37	\pm	0.46	&	0.89	\pm	0.43	&	0.73	\pm	0.36	&	5.89	&	38.24	&		\\
\rm NGC\,3228	&	2.9	&	4.36	\pm	3.25	&	3.0	\pm	0.41	&	2.34	\pm	0.48	&	0.69	\pm	0.52	&	0.54	\pm	0.41	&	0.32	&	24.11	&		\\
\rm NGC\,3532	&	3.4	&	12.32	\pm	4.52	&	9.7	\pm	0.53	&	5.06	\pm	0.6	&	0.79	\pm	0.29	&	0.41	\pm	0.16	&	1.21	&	11.37	&	\rm III	\\
\rm NGC\,6405	&	4.4	&	9.87	\pm	7.42	&	4.6	\pm	0.69	&	3.26	\pm	0.77	&	0.47	\pm	0.36	&	0.33	\pm	0.26	&	3.82	&	12.03	&		\\
\rm NGC\,6475	&	2.0	&	9.8	\pm	1.52	&	8.0	\pm	0.23	&	4.73	\pm	0.29	&	0.82	\pm	0.13	&	0.48	\pm	0.08	&	2.05	&	5.49	&	\rm I	\\
\rm NGC\,7058	&	2.7	&	4.45	\pm	2.81	&	3.24	\pm	0.38	&	2.43	\pm	0.45	&	0.73	\pm	0.47	&	0.55	\pm	0.36	&	19.24	&	72.08	&		\\
\rm NGC\,7092	&	2.3	&	6.37	\pm	2.01	&	5.11	\pm	0.3	&	4.01	\pm	0.35	&	0.8	\pm	0.26	&	0.63	\pm	0.21	&	5.5	&	82.13	&		\\
\rm Pleiades$^b$	&	0.9	&	7.9	\pm	0.23	&	7.06	\pm	0.07	&	4.02	\pm	0.07	&	0.89	\pm	0.03	&	0.51	\pm	0.02	&	17.79	&	10.35	&	\rm I	\\
\rm Praesepe	&	1.6	&	7.37	\pm	0.97	&	5.32	\pm	0.15	&	4.95	\pm	0.2	&	0.72	\pm	0.1	&	0.67	\pm	0.09	&	11.14	&	43.6	&		\\
\rm Roslund\,5	&	4.1	&	5.67	\pm	6.39	&	4.51	\pm	0.64	&	4.24	\pm	0.71	&	0.8	\pm	0.9	&	0.75	\pm	0.85	&	9.21	&	33.44	&		\\
\rm RSG\,7	&	2.4	&	4.74	\pm	2.27	&	3.8	\pm	0.34	&	2.36	\pm	0.4	&	0.8	\pm	0.39	&	0.5	\pm	0.25	&	9.39	&	33.79	&	\rm I	\\
\rm RSG\,8	&	4.8	&	8.6	\pm	8.99	&	7.01	\pm	0.76	&	6.1	\pm	0.85	&	0.82	\pm	0.86	&	0.71	\pm	0.75	&	20.71	&	44.6	&		\\
\rm Stephenson\,1	&	2.6	&	5.72	\pm	2.57	&	5.23	\pm	0.36	&	4.42	\pm	0.41	&	0.91	\pm	0.42	&	0.77	\pm	0.35	&	26.02	&	53.14	&		\\
\rm Stock\,1	&	2.4	&	5.41	\pm	2.23	&	4.83	\pm	0.33	&	3.16	\pm	0.39	&	0.89	\pm	0.37	&	0.58	\pm	0.25	&	2.66	&	81.57	&	\rm I	\\
\rm Stock\,12	&	2.7	&	6.5	\pm	2.73	&	4.62	\pm	0.38	&	2.51	\pm	0.44	&	0.71	\pm	0.3	&	0.39	\pm	0.18	&	10.3	&	88.02	&	\rm III	\\
\rm Stock\,23	&	4.0	&	7.52	\pm	6.15	&	3.83	\pm	0.65	&	2.78	\pm	0.71	&	0.51	\pm	0.43	&	0.37	\pm	0.32	&	33.59	&	57.15	&		\\
\rm UBC\,7$^b$	&	2.0	&	6.95	\pm	1.59	&	4.66	\pm	0.26	&	3.83	\pm	0.28	&	0.67	\pm	0.16	&	0.55	\pm	0.13	&	12.48	&	26.29	&		\\
\rm UBC\,19	&	3.5	&	3.94	\pm	4.87	&	3.4	\pm	0.56	&	2.23	\pm	0.62	&	0.86	\pm	1.07	&	0.57	\pm	0.72	&	14.83	&	34.2	&	\rm I	\\
\rm UBC\,31	&	3.2	&	6.31	\pm	4.06	&	5.74	\pm	0.5	&	4.67	\pm	0.57	&	0.91	\pm	0.59	&	0.74	\pm	0.48	&	6.78	&	33.83	&		\\
\rm UBC\,31\,gp1$^a$	&	3.4	&	6.07	\pm	4.56	&	4.13	\pm	0.53	&	3.07	\pm	0.61	&	0.68	\pm	0.52	&	0.51	\pm	0.39	&	1.68	&	11.99	&		\\
\rm UBC\,31\,gp2$^a$	&	3.0	&	6.99	\pm	3.6	&	6.1	\pm	0.46	&	4.6	\pm	0.53	&	0.87	\pm	0.45	&	0.66	\pm	0.35	&	6.1	&	27.11	&	\rm III	\\
\rm UPK\,82	&	3.7	&	4.58	\pm	5.24	&	3.49	\pm	0.57	&	2.65	\pm	0.66	&	0.76	\pm	0.88	&	0.58	\pm	0.68	&	32.52	&	72.08	&		\\
\enddata
\tablecomments{\\
	$^a$ New hierarchical groups identified in this work. \\
	$^b$ Clusters members taken from \citet{pang2021b,li2021}.\\
    Column 2, $\sigma_{dis}$, shows the uncertainty of corrected distance of each cluster (Section~\ref{sec:dis_corr}). In columns 3--5 are the semi-major ($a$), semi-intermediate ($b$) and semi-minor ($c$) axes of the fitted ellipsoid for each star cluster. $\theta$ is the angle between the direction of $a$ and the Galactic plane and $\phi$ is the angle between the direction of $a$ and the line-of-sight. Column 10 show the type of fitted ellipsoid for the bound region with Type~I stands for ``oblate'', II for ``prolate'', and III for ``triaxial''.
    }
\end{deluxetable*}

\section{3D Morphology Discussion}\label{sec:3ddis}

The morphology of an open cluster evolves as it ages. It changes from inherent filamentary or fractal substructures to acquired halo or tidal tails. The variation of spatial distribution reflects different dynamical processes dominating the current cluster evolution. Inside the tidal radius, the influence of the cluster's gravitational potential is competing with Galactic tides. It is intriguing to investigate the similarity and differences in the morphology of young clusters' bound regions to explore the early dynamical evolution of clusters.  

\begin{figure*}[ht!]
\centering
\includegraphics[scale=0.8]{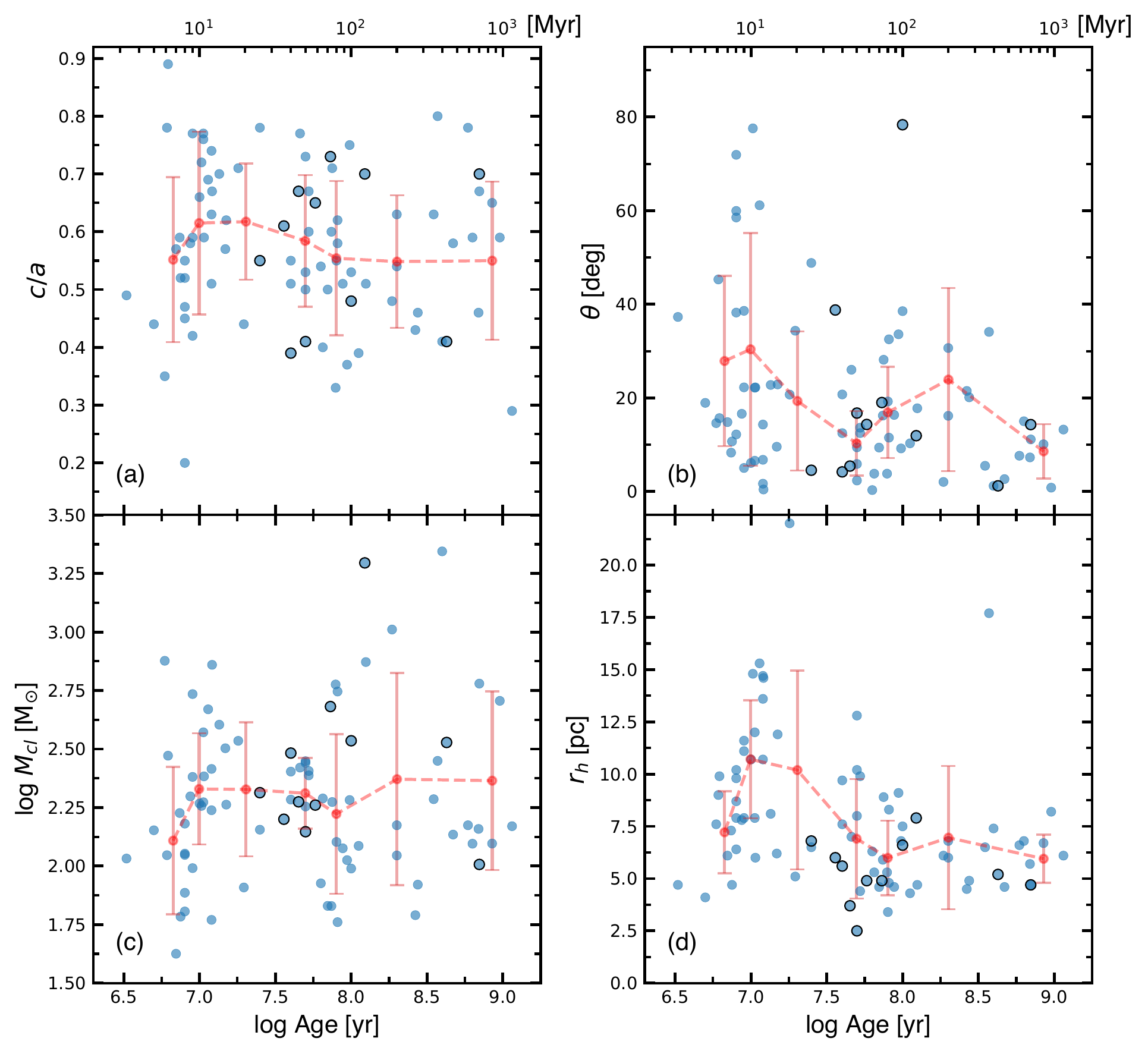}
\caption{
    Scatter diagrams with error bar of the relationship between four parameters and the logarithm of cluster age. (a) Relationship of the age with axis ratio $c/a$. (b) Relationship with $\theta$, the angle between the cluster elongation and the Galactic Plane. (c) Relationship with the logarithm of cluster mass $M_{cl}$. (d) Relationship with half-mass radius $r_{\rm h}$. The blue dots are the values for the 84 clusters (Group\,X excluded). The red dot and error bar are the mean and standard deviation in each bin. To ensure a statistical significance, we keep a constant number of 12 clusters in each bin. Parameters of 12 open clusters (highlighted with black circles) are taken from paper~I.}
\label{fig:relation-age}
\end{figure*}

The relationship between clusters' spatial morphology and other properties can be linked to their evolution in space and time. We present the correlation between cluster age and other parameters in Figure~\ref{fig:relation-age}. 
The mean ratio $c/a$ seems to increase as the cluster grows older until 30\,Myr (panel (a)), though a large scatter is observed. 
Most clusters younger than $\sim$30\,Myr are filamentary or fractal. These young clusters are elongated in a direction with a random inclination to the Galactic plane (panel (b)). The mean value of $\theta$ tends to decrease from $\sim$30\,degree at $\sim$10\,Myr to $\sim$10--15\,degree in $\sim$30\,Myr (panel (b)).

Many filamentary young clusters (groups) with small $c/a$ are dispersing (Section~\ref{alessi20}) so that they were not able to grow older than 30\,Myr. At the same time, other groups in the hierarchical clustering region follow the infalling flow and merger to form a denser cluster (Section~\ref{ic348} and Section~\ref{lp2442}). During this merging event, the primordial filamentary morphology is significantly modified, and produces a more spherical morphology by the dynamical mixing. The inclination of the elongation direction of the cluster to the Milky Way disk, $\theta$, as a result, is significantly reduced.


The apparent increasing trend of the mean value of $c/a$, $M_{cl}$, and $r_h$  for the young clusters (groups) can be explained by filamentary group dissolution and group mergers in the hierarchical formation framework. Though the age correlation has a large scatter, this overall picture derived from it reaches an agreement with our morphological and kinematic analysis of clusters in previous sections (Sections~\ref{sec:3D_solar}, \ref{sec:kin_feature}, and~\ref{sec:hiearchy}).

The mean value of $c/a$ declines after $\sim$30\,Myr, and stays at $\sim0.5$ for age above $\sim$100\,Myr (panel (a)). The inclination angle to the Galactic plane, $\theta$, is generally smaller in old clusters (panel (b)). 
As Galactic tide becomes increasingly dominant in older clusters, clusters are elongated again (smaller $c/a$) with the development of tidal tails. This disruption process should go along with mass loss; however, we do not observe a significant decline of cluster mass in old clusters (panel (c)). This inconsistency shows that only the most massive and dense clusters can survive long enough under the influence of violent Galactic force. The decrease of $r_{\rm h}$ in older clusters is attributed to the two-body relaxation that generates more compact clusters \citep{heggie2003}, consistent with the finding of $r_c$ in Section~\ref{sec:radial_density}.

Observations have found that clusters younger than $\sim$30\,Myr are mostly moving around the Galactic center in circular orbits \citep{tarricq2021}. They follow much lower vertical velocities than the old clusters \citep{soubiran_open_2018,kounkel2020}. It seems that they are moving slowly in the dense star formation region in the spiral arms (Figure~\ref{fig:xyz_2D}) and facing disruption and merger at the same time. On the other hand, old clusters run faster along the $Z$ direction and manage to locate further away from the disk, which alleviates their disruption rate.   


\section{Summary}\label{sec:summary}

We obtain membership for 65 open clusters and stellar groups in the solar neighborhood via \texttt{StarGO} based on Gaia EDR3 data. Ten groups of stars in four Hierarchical clustering regions (hosting 29 groups) are newly discovered. Those newly discovered groups in different sky regions are: (1) Alessi\,20 gp\,1 and Alessi\,20 isl\,1 in the Alessi\,20 region; (2) LP\,2373 gp\,1 in the LP\,2373 region; (3) UBC\,31 gp\,1 and UBC\,31 gp\,2 in the IC\,348 region; (4) LP\,2442 gp\,1--5 in the LP\,2442 region.

We verify the robustness of cluster membership by calculating the recovery rate of members from the simulated observation data set (combination of $N$-body simulated clusters and Gaia EDR3 mock field stars). The overall morphology of clusters can be strongly affected by a contamination rate $>$20\%. An artificial ``halo''-like substructure will appear when the contamination rate rises above 20\%. The bound members (stars within tidal radius), however, unlike those unbound, are more stable in their membership. The recovery rate of bound members remains above 99\% despite the overall contamination changing from 10\% to 25\%.

Besides the 65 clusters and groups, we adopt twelve clusters from \citetalias{pang2021a}, seven open clusters and groups in Vela~OB2 from \citet{pang2021b}, and the Pleiades from \citep{li2021}. The morphological parameters of ellipsoidal fitting a total of 72 open clusters (groups) are being computed for the first time by this work. Based on the morphological parameters derived from ellipsoidal fitting, the 3D morphology of the bound region of 14 clusters resemble oblate spheroid, three prolate spheroid, and seven triaxial ellipsoids.

We adopt the morphological parameters of twelve clusters from \citetalias{pang2021a}. In total, 85 clusters are analyzed for the 3D morphology investigation. We classify the morphological substructures detected in these 85 clusters into four types: (1) the Filamentary (f1) type, for clusters younger than 100\,Myr and having uni-directional elongated substructures; (2) the Fractal (f2) type, for clusters that are as young as the filamentary type but with multi-directional and fluffy spatial substructures; (3) the Halo (h) type, for clusters older than 100\,Myr and harbor a diffuse halo around the massive core; (4) the Tidal-tail (t) type, for old clusters ($>$100\,Myr) with extended tidal tails.

The radial density profile of 82 clusters is fitted with the EFF profile \citep{elson1987}. Group X, Alessi\,20 gp1 and UBC\,31 gp2 are excluded in the fitting owing to irregular shapes. As a result of tidal disruption, t-type clusters have the smallest size of cluster core and have the most shallow radial density profile. Although clusters in the h-type are as old as those in the t-type, their high-mass and high-density nature slow down their disruption process. 

Kinematic substructures are identified among all four types of morphological substructures. Clusters of f1-type have elongated kinematic substructures in the proper motion space, similar to the disrupted cluster Group\,X. This elongated feature seen in the velocity space for the f1-type clusters shows their unstable and short-lived dynamical state. Kinematic tail extended from the dense core in the proper motion diagram is most prominent in the t-type clusters, especially the Pleiades.

Morphology and kinematics of hierarchical clustering groups are highly complex. They form extended filament networks like in the Alessi\,20, IC\,348, and LP\,2373 regions. Global ``orthogonal'' expansion seems to disperse the filaments in Alessi\,20 and LP\,2373. Tangential in-falling motion along filament direction is detected in UBC\,31, UBC\,31 gp\,1, and UBC\,31 gp\,2 in the IC\,348 region. Hierarchical subgroups in the LP\,2442 region are well mixed in positions and are distinctive in proper motion space. The complexity of the 3D motions in LP\,2442 gp\,1--5 might be attributed to an ongoing merger event.

Correlations between cluster age and the three structural parameters, the axis ratio $c/a$,  the inclination of cluster elongated direction $a$ to the Galactic plane $\theta$, cluster mass $M_{cl}$, and half-mass radius $r_{\rm h}$, are observed. The mean value of $c/a$, $M_{cl}$, and $r_{\rm h}$ all follow an increasing trend with larger cluster ages when below $\sim$30\,Myr. The mean $\theta$, on the contrary, shows decreasing trend before $\sim$30\,Myr. These signs imply two possible dynamical processes that young clusters (groups) may experience before 30\,Myr. (1) Young stellar groups with low SFE (mostly along filaments in the molecular clouds) quickly dissolve and become unbound. (2) Filaments at high SFE, at the location of the hub or major filaments in the molecular cloud, have a chance to merge and form a dense cluster. 
\clearpage

\begin{acknowledgments}
We thank the anonymous referee for advice to improve the paper.
We are also grateful for the IT support of Lowell Observatory, and thank Xianhao Ye and Chaojie Hao for helpful discussion. This work is supported by the grant of National Natural Science Foundation of China, No: 12173029. 
X.Y.P. is grateful to the financial support of the Research Development Fund of Xi'an Jiaotong-Liverpool University (RDF-18--02--32). X.Y.P. gave thanks to the funding of Natural Science Foundation of Jiangsu Province, No: BK20200252. 
M.B.N.K. is grateful for the support from the Continuous Support Fund (grant RDF-SP-93) of Xi’an Jiaotong-Liverpool University. 

This work made use of data from the European Space Agency (ESA) mission {\it Gaia} 
(\url{https://www.cosmos.esa.int/gaia}), processed by the {\it Gaia} Data Processing 
and Analysis Consortium (DPAC, \url{https://www.cosmos.esa.int/web/gaia/dpac/consortium}). This study also made use of 
the SIMBAD database and the VizieR catalogue access tool, both operated at CDS, Strasbourg, France.
\end{acknowledgments}

\software{  \texttt{Astropy} \citep{astropy2013,astropy2018}, 
            \texttt{SciPy} \citep{millman2011},
            \texttt{numpy} \citep{harris2020array},
            \texttt{galpy} \citep{Bovy2015},
            \texttt{TOPCAT} \citep{taylor2005}, and 
            \texttt{StarGO} \citep{yuan2018},
            \texttt{Plotly} \citep{plotly},
            \texttt{petar} \citep{Wang2020b}
}

\clearpage
\bibliography{main}
\bibliographystyle{aasjournal}

\end{document}